\documentclass[aps,prl,superscriptaddress,two column,preprintnumbers,notitlepage]{revtex4-1}
\usepackage{latexsym}
\usepackage{subcaption}
\usepackage{caption}

\usepackage[dvips]{color}
\usepackage{graphicx}
\usepackage{amsmath}
\usepackage{amssymb}
\usepackage{gensymb}
\usepackage{enumerate}
\usepackage{hyperref}
\usepackage{mathrsfs}
\usepackage[usenames,dvipsnames]{xcolor}
\usepackage{float}
\def\mr{\mathrm}

\begin{document}

\title{Excess entropy and breakdown of semiclassical description of thermoelectricity in twisted bilayer graphene close to half filling}

\author{Bhaskar Ghawri}
\email{gbhaskar@iisc.ac.in}
\author{Phanibhusan S. Mahapatra}
\email{phanis@iisc.ac.in}
\author{Shinjan Mandal}
\author{Aditya Jayaraman}
\affiliation{Department of Physics, Indian Institute of Science, Bangalore, 560012, India}
\author{Manjari Garg}
\affiliation{Department of Instrumentation and Applied Physics, Indian Institute of Science, Bangalore, 560012, India}
\author{K. Watanabe}
\affiliation{Research Center for Functional Materials, National Institute for Materials Science, Namiki 1-1, Tsukuba, Ibaraki 305-0044, Japan}
\author{ T. Taniguchi}
\affiliation{International Center for Materials Nanoarchitectonics, National Institute for Materials Science, Namiki 1-1, Tsukuba, Ibaraki 305-0044, Japan}
\author{H. R. Krishnamurthy}
\author{Manish Jain}
\affiliation{Department of Physics, Indian Institute of Science, Bangalore, 560012, India}
\author{Sumilan Banerjee}
\affiliation{Department of Physics, Indian Institute of Science, Bangalore, 560012, India}
\author{U. Chandni}
\affiliation{Department of Instrumentation and Applied Physics, Indian Institute of Science, Bangalore, 560012, India}
\author{Arindam Ghosh}
\email{arindam@iisc.ac.in}
\affiliation{Department of Physics, Indian Institute of Science, Bangalore, 560012, India}
\affiliation{Centre for Nano Science and Engineering, Indian Institute of Science, Bangalore 560 012, India}

\maketitle

\textbf{
In moir\'{e} systems with twisted bilayer graphene (tBLG), the amplification of Coulomb correlation effects at low twist angles ($\theta$) is a result of nearly flat low-energy electronic bands \citep{trambly2010localization,bistritzer2011moire} and divergent density of states (DOS) at van Hove singularities (vHS) \citep{Yuan2019}. This not only causes superconductivity~\cite{cao2018unconventional}, Mott insulating states~\cite{cao2018correlated}, and quantum anomalous Hall effect~\cite{sharpe2019emergent} close to the critical (or magic) angle $\theta = \theta_\mr{c} \approx 1.1^\circ$, but also unconventional metallic states that are claimed to exhibit non-Fermi liquid (NFL) excitations~\cite{cao2020strange}. However, unlike superconductivity and the correlation-induced gap in the DOS, unambiguous signatures of NFL effects in the metallic state remain experimentally elusive. Here we report simultaneous measurement of electrical resistivity ($\rho$) and thermoelectric power ($S$) in tBLG at $\theta \approx 1.6^\circ$. We observe an emergent violation of the semiclassical Mott relation in the form of excess $S$ close to half-filling. The excess $S$ ($\approx 2$~$\mu$V/K at low temperature $T \sim 10$~K) persists up to $\approx 40$~K, and is accompanied by metallic $T$-linear $\rho$ with transport scattering rate ($\tau^{-1}$) of near-Planckian magnitude $\tau^{-1} \sim k_\mr{B}T/\hbar$ \citep{bruin2013similarity}. The combination of non-trivial electrical transport and violation of Mott relation provides compelling evidence of NFL physics intrinsic to tBLG, at small twist angle and half-filling.}

The phenomenological similarities between superconductivty in tBLG and that in high-$T_\mr{c}$ cuprates~\cite{lee2006doping} leads one to question the validity of Landau quasiparticle in the former for twist angles near  $\theta = \theta_\mr{c}$. Even for temperatures $T > T_\mr{c}$, where $T_\mr{c}$ is the superconducting transition temperature, a linear $T$-dependence of the resistivity ($\rho$) near half-filling (or equivalently, band filling factor $\nu = \pm 2$) of the four-fold spin-valley degenerate conduction and valence bands seems to indicate the absence of well-defined quasiparticle spectrum~\cite{cao2020strange}. On the contrary, persistence of the linearity in $\rho$ for $\theta$ well away from $\theta_\mr{c}$, {\it e.g.} for $\theta \sim 1.5 - 2^\circ$, led other theoretical~\cite{wu2019phonon} and experimental~\cite{polshyn2019large} investigations to view the tBLG in this regime as a two dimensional, weakly (or non-) interacting metal with largely reduced Bloch-Gr$\mr{\ddot u}$neisen temperature ($T_\mr{BG}$). In scanning tunneling microscopy \citep{kerelsky2019maximized,jiang2019charge,Lin2019Magnetism}  experiments, although possibility of an interaction-driven magnetic order has been claimed close to the vHS for $\theta \approx 1.6\degree$, the spontaneous breaking of $C_6$ lattice symmetry to nematic orbital order has not been observed for $\theta > \theta_\mr{c}$. Thus away from $\theta_\mr{c}$, the impact of electronic correlation at small $\theta \lesssim 2^\circ$ remains uncertain, even though the renormalization of the Fermi velocity and localization at AA sites are still significant~\citep{trambly2010localization,bistritzer2011moire}.  

\begin{figure*}[bth]
  \includegraphics[width=1.0\textwidth]{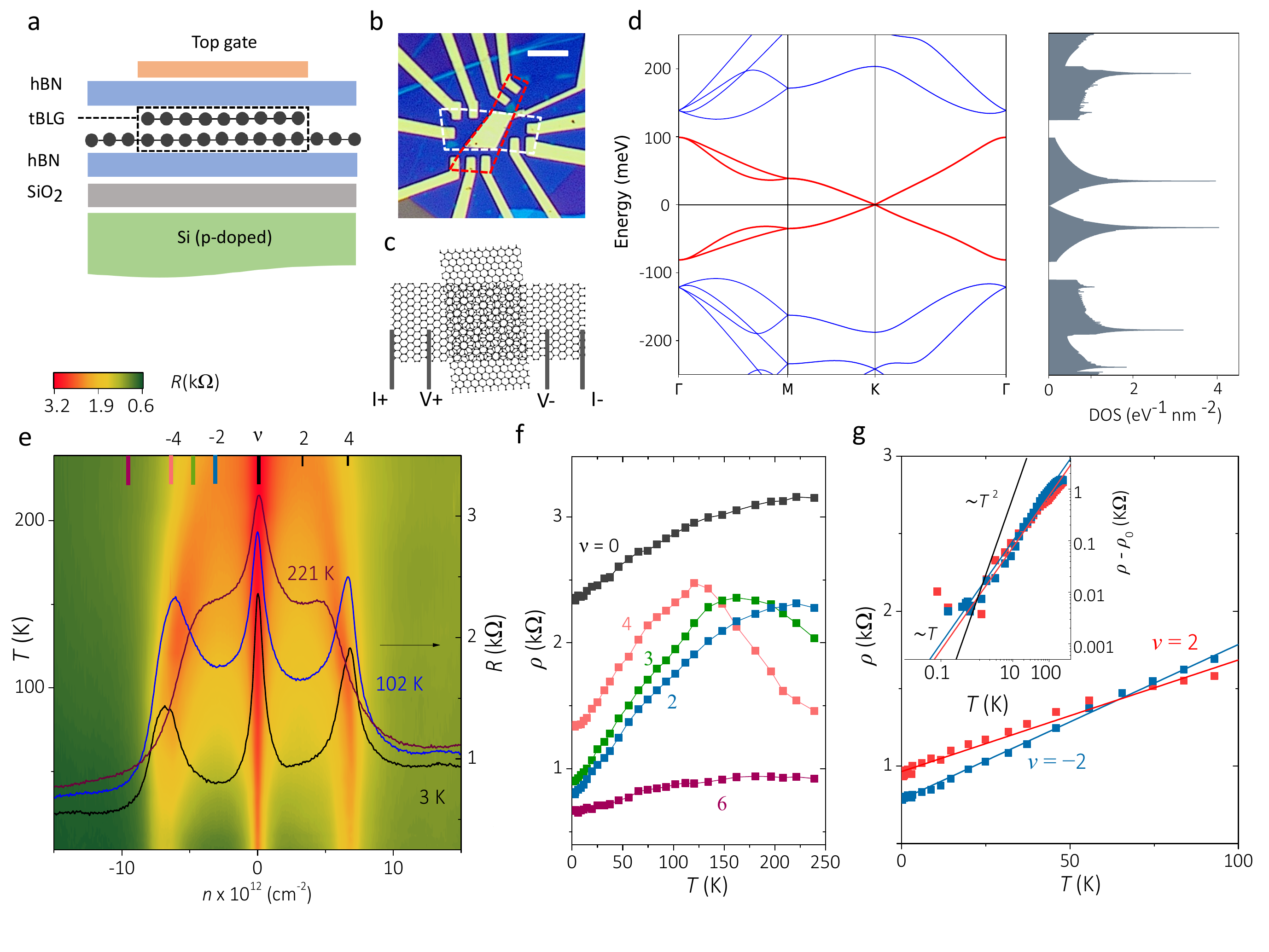}
  \captionsetup{justification=raggedright,singlelinecheck=false}
\caption{\textbf{ Device structure, electronic band structure and electrical transport}. (a) Schematic of the cross-sectional view of the device showing the constituent layers. (b) Optical image of the device. The scale bar represents a length of $5$~$\mu$m. (c) Schematic showing the contact configuration for the four-terminal measurement of resistance $R$. (d) Electronic band structure and density of states (DOS) of tBLG ($\theta = 1.6 \degree$) calculated using tight binding model. The bands shown in red are the low energy active bands. (e) Surface plot of $R$ as a function of $T$ and $n$. The solid curves show density variation of $R$ on the right axis at three representative temperatures $3$~K, $102$~K and $221$~K, respectively. (f) Resistivity $\rho$ as a function of $T$ for selected values of $\nu$ which are marked with vertical dashes in (e). (g) $T$-dependence of $\rho$ at $\nu=\pm 2$ in $T$-linear regime. Solid lines show the $T$-linear fit to the data. The inset shows $T$-dependence of $\rho-\rho_{0}$ at $\nu=\pm 2$ in logarithmic scale. Solid lines show $T$-linear and $T^2$ dependences, respectively.}
\end{figure*}

Here we have carried out simultaneous electrical and thermoelectric measurements in tBLG misoriented at $\theta \approx 1.6^\circ$. The dependence on $T$ and on the carrier density ($n$) of the thermoelectric power ($S$), or the Seebeck coefficient, is used as an independent and sensitive probe of the correlation effects. Thermoelectric power is often interpreted as a thermodynamic entity that represents the entropy carried by each charge carrier.  Within the degenerate  quasiparticle description in the Boltzmann transport regime ($T \ll T_\mr{F}$, where $T_\mr{F}$ is the Fermi temperature), $S$ is related to the resistance ($R$) through the semiclassical Mott relation (SMR),
\begin{equation}
\label{Semiclassical Mott relation}
S_\mr{Mott}=\frac{\pi^2 k_\mr{B}^2 T}{3|e|} \frac{\mr{dln}R(E)}{\mr{d}E}\bigg\vert_{E_\mr{F}},
\end{equation}
where $R$, $e$ and $E_\mr{F}$ are energy-dependent resistance, electronic charge and Fermi energy, respectively. Eq.~\ref{Semiclassical Mott relation} is valid under the assumption that scattering is elastic and isotropic throughout the Fermi surface \textit{i.e.} transport liftime only depends on the energy of the charge carriers. Remarkably, this simple assumption of isotropic scattering remains valid in a wide variety of systems, such as disordered metals/semiconductors \citep{rowe2017materials,behnia2015fundamentals}, organic materials \citep{watanabe2019validity}, monolayer graphene~\cite{zuev2009thermoelectric} and topological insulators~\cite{kim2014ambipolar}. The SMR effectively arises from the quasiparticles carrying heat and charge under identical constraints, imposed by the momentum conservation. Thus, the validity of SMR in Eq.~\ref{Semiclassical Mott relation} provides a definitive probe into the scattering mechanisms and energy distribution of the charge carriers near the Fermi surface, which breaks down when strong correlation effects become important \cite{arsenijevic2013signatures,behnia2015fundamentals}.

\begin{figure*}[t]
  \includegraphics[width=1.0\textwidth]{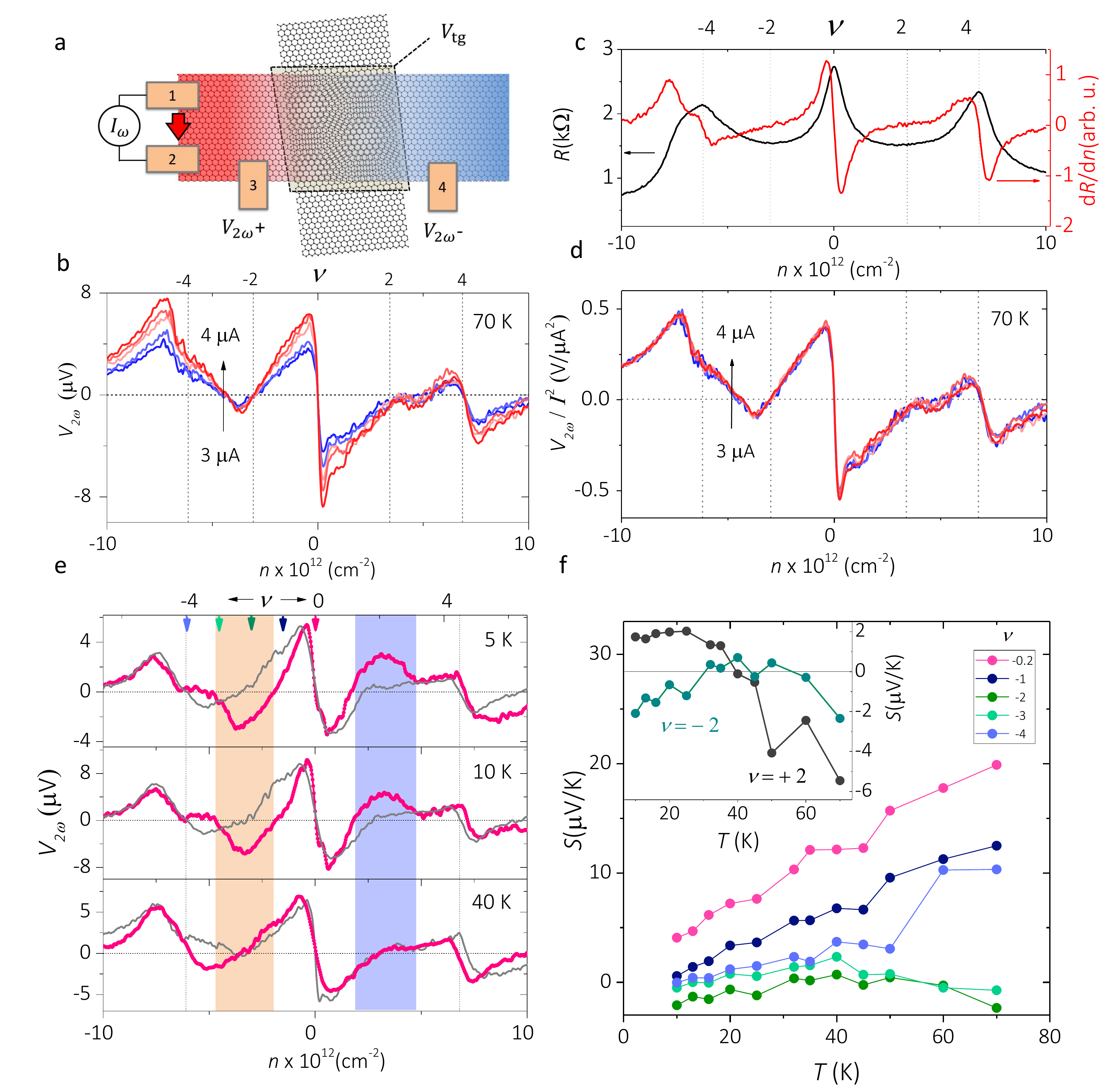}
  \captionsetup{justification=raggedright,singlelinecheck=false}
\caption{\textbf{ Thermoelectric transport in twisted bilayer graphene:} (a) In-plane heating and measurement schematic for thermo-voltage $V_{2\omega}$. (b) $V_{2\omega}$ as a function of $n$ for different heating currents ($3-4$~$\mu$A) at $70$~K. (c),(d) Simultaneously measured $R$ and $V_{2\omega}$ normalized with $I_{\omega}^2$ at $70$~K. The right axis in (c) shows the numerically calculated $\mr{d}R/ \mr{d}n$. (e) Comparison between the measured $V_{2\omega}$ (pink lines) and that calculated (Grey line) from the semiclassical Mott relation (Eq.~\ref{Mott relation}) at three representative temperatures. $\Delta T$ is obtained as a fitting parameter to match SMR with the experimental $V_{2\omega}$ at CNP. (f) Temperature dependence of $S$ at various band filling factors which are marked with arrows in (d). The inset shows the $T$ dependence of $S$ at $\nu = \pm 2$.} 
\end{figure*}

The tBLG device for our experiment was created using standard van der Waals stacking \citep{mahapatra2017seebeck}, which consists of two graphene layers aligned at $60^{\circ} + \theta$, thus $\theta$ being the effective twist angle~\citep{mahapatra2019misorientation}, and encapsulated within two sheets of hexagonal boron nitride (hBN) (see schematic shown in Fig.~1a). The moir\'{e} super-lattice is formed at the overlap region ($\approx 5$~$\mu$m~$\times$~$6$~$\mu$m), and the monolayer branches of graphene on four sides act as electrical leads. The device micrograph is shown in Fig.~1b. A local top-gate tunes $n$ of the overlap region, while the global, doped silicon backgate is usually kept at a large value ($\approx -35$~V) to minimize the contact resistance and thermovoltage contributions from outside the overlap region. Both electrical and thermovoltage measurements show consistent results across different thermal cycles (see supplementary information, SI, section III). Fig.~1e shows the resistance $R$ measured in the four-terminal configuration (Fig.~1c) across the overlap region, as a function of $n$ (by varying the top gate voltage $V_\mr{tg}$) and $T$. We observe three resistance peaks (right axis in Fig.~1e), located at the charge neutrality point (CNP) and at $ n \approx \pm 6.4 \times 10^{12}$~cm$^{-2}$ for $T\lesssim 100$~K, where the latter correspond to full filling of the lowest band of the tBLG super-lattice ({\it i.e.} $\nu = \pm 4$)~\citep{cao2016superlattice,kim2016charge}. This was independently verified from the evolution of Landau fans in $R$ originating from the $\nu = \pm 4$ in perpendicular magnetic field (see SI, section IV). From the corresponding moir\'{e} period, we estimate the twist angle $\theta \approx 1.6 \degree$. A tight binding calculation for the electronic band structure and the corresponding DOS for $\theta = 1.6^{\circ}$ are shown in Fig.~1d. The active low-energy bands, shown in red, have a width $W \sim 180$~meV, while the (indirect) gap between the active and higher energy bands is $\Delta_\mr{s} \sim 27$~meV, for both electron and hole sides (see Methods and SI, section V for more details on the band structure calculations).

\begin{figure}[t]
  \includegraphics[clip,width=9cm]{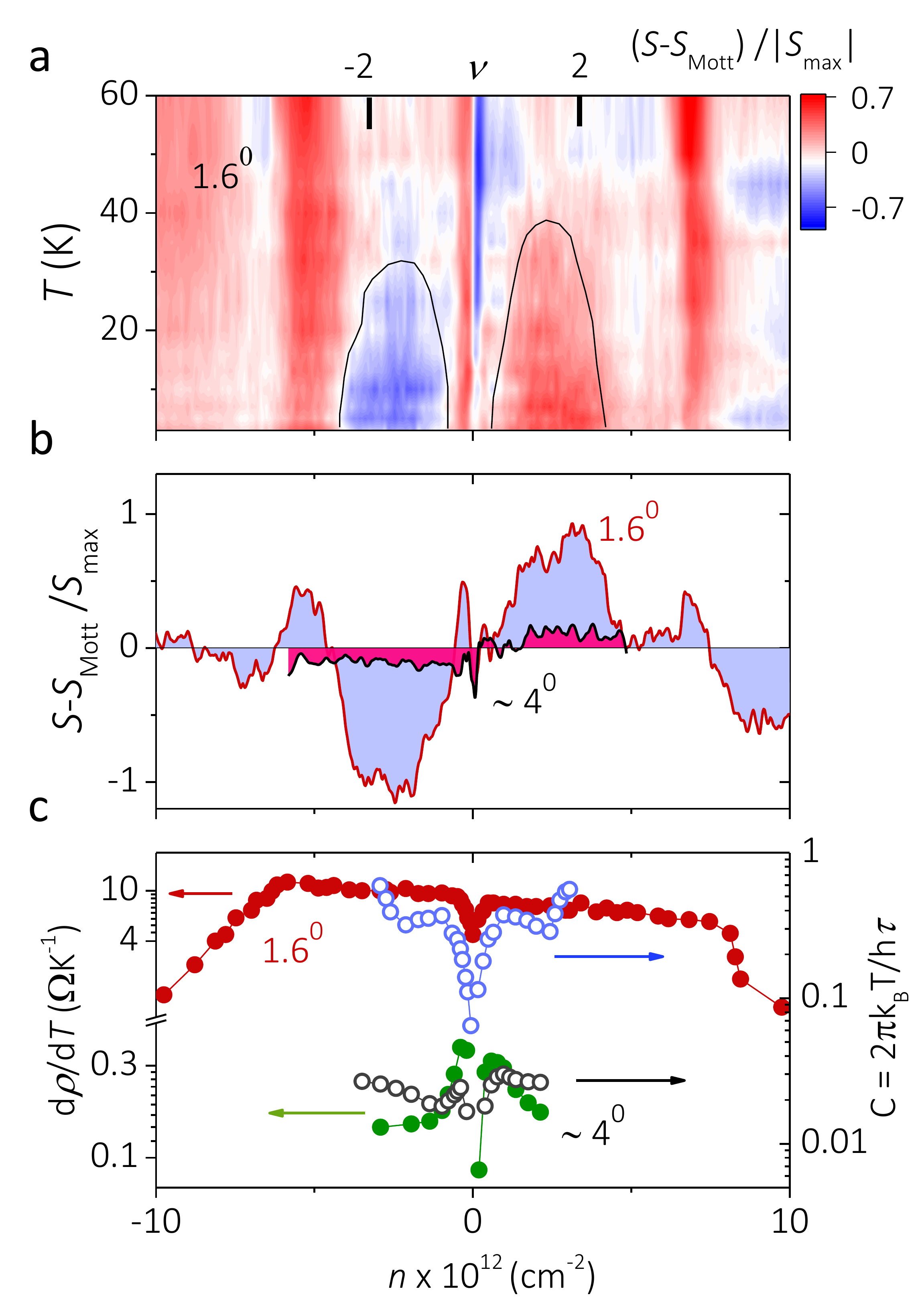}
  \captionsetup{justification=raggedright,singlelinecheck=false}
\caption{\textbf{Breakdown of semiclassical Mott relation and scattering rate :} (a) Surface plot of $(S-S_\mr{Mott})/S_\mr{max}$, as a function of $T$ and $n$ for $\theta \approx 1.6 \degree$. (b) ($(S-S_\mr{Mott})/S_\mr{max}$) at $5$~K for $\theta \approx 1.6 \degree$ and $ \theta \sim 4 \degree$. (c) $\mr{d}\rho/ \mr{d}T$ extracted in the $T$-linear regime at different $n$ for $\theta \approx 1.6 \degree$ (red circles) and  $\theta \sim 4 \degree$(green circles). The right axis shows dimensionless pre-factor $C$ of the scattering rate $\Gamma = Ck_\mr{B}T/\hbar$ for (c) $\theta \approx 1.6 \degree$ (open blue circles) and (d) $\theta \sim 4 \degree$ (open black circles).}
\end{figure}

The apparent shift in the resistance peak position at $\nu = \pm 4$ for $T \gtrsim 150$~K in Fig.~1e is due to a metal to insulator-like crossover in $\rho$ ($R/\square$) at finite doping (Fig.~1f). Focusing within the active band ({\it i.e.} $-4\leq\nu\leq4$), we find that $\rho$ is insulating for $T \gtrsim T_\mr{H}$, where $T_\mr{H} \sim 100 - 200$~K is a doping-dependent characteristic temperature (see SI, section VI), but becomes metallic at $T \lesssim T_\mr{H}$ and remains so down to the lowest experimental temperature ($\approx 100$~mK). The absence of insulating state at $\nu = 0$ (Dirac point) is likely to be a combination of inhomogeneity and relatively weak {\it e-e} interactions that fails to lift the $C_3$ or $C_2\mathcal{T}$ symmetries \citep{lu2019superconductors}. The insulating transition at $T > T_\mr{H}$ has been previously attributed to thermally activated transport of charge carriers to the dispersive higher energy bands~\citep{polshyn2019large}, which seems to be case here too as $T_\mr{H} \approx 100$~K is lowest for $\nu = \pm4$. At $T \lesssim T_\mr{H}$, we find $\rho$ to vary as $\rho=\rho_{0}+AT$ where $\rho_0$ is the residual resistivity. The order of $A$ ($\sim 10$~$\Omega$/K) and $\rho$ ($\sim 1 - 3$~k$\Omega$), are both consistent with the earlier transport measurements in tBLG at $\theta \approx 1.6^\circ$~\cite{polshyn2019large}. The metallicity was observed at {\it all} fillings including, unexpectedly, at the super-lattice gap ($\nu = \pm4$). While this is not understood at the moment, we cannot rule out the possibility of a correlated metallic state due to competing interaction energy and relatively small $\Delta_\mr{s}$ \cite{bag2019correlation}. The $T$-linearity of $\rho$ is most pronounced at $\nu = \pm2$ with $T_\mr{H} \gtrsim 250$~K (Fig.~1g). As further emphasized in the inset that shows $\rho-\rho_0$ vs. $T$ in logarithmic scale, we find clear departure from  $\rho \sim T^2$ dependence associated with electron-electron scattering, or the $\rho\propto T^4$ behavior, expected due to electron-acoustic phonon scattering at $T \ll T_\mr{BG}$~\citep{efetov2010controlling}.

\begin{figure*}[t]
  \includegraphics[width=1.0\textwidth]{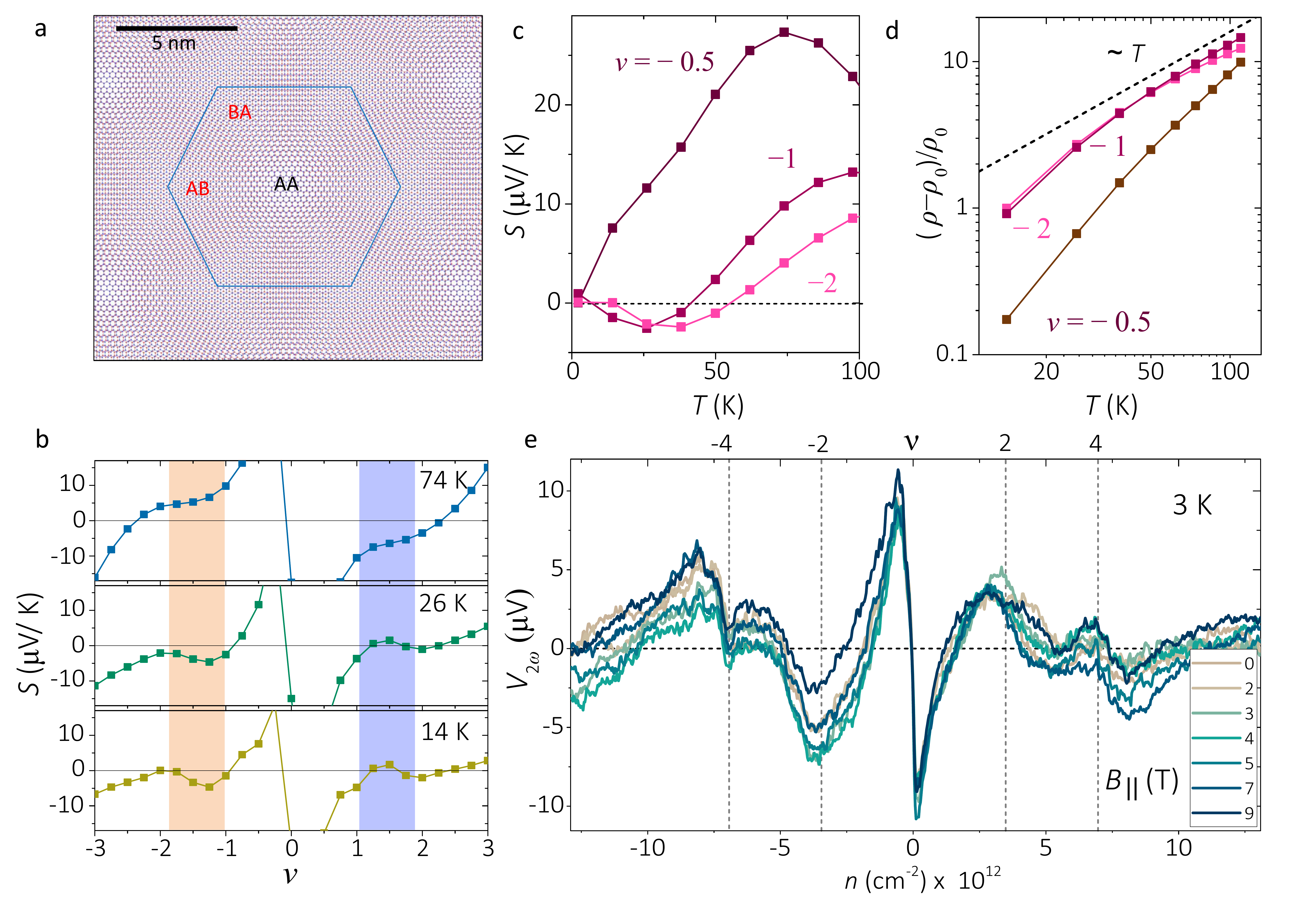}
  \captionsetup{justification=raggedright,singlelinecheck=false}
\caption{{\bf Dynamic mean field theory (DMFT) results and magneto-Seebeck measurements:} (a) Schematic of a tBLG moir\'{e} super-lattice at $1.6\degree$. AA-staked regions are surrounded by AB/BA stacked regions. (b) Seebeck coefficient $S$ computed in DMFT with $U=38$ meV as a function of filling $\nu$ for the four lowest bands at three temperatures $T=14$~K, $26$~K and ~$74$ K, respectively. (c) Computed $S$ as a function of temperature for  fillings $\nu=-2$,~$-1$~ and $-0.5$. The Seebeck coefficient changes from positive to negative sign over an intermediate temperature range. (d) Normalized  resistivity  $\rho$ computed as a function of $T$ for fillings  $\nu=-2$,~$-1$~ and $-0.5$. The dashed line shows $T$-linear dependence. (e) $V_{2\omega}$ as a function of $n$ for different magnetic fields applied parallel to the plane of the tBLG.} 
\end{figure*}

To complement electrical transport, we then perform thermoelectric measurements in the same device. The measurements are schematically explained in Fig.~2a. Briefly, a sinusoidal current ($I_{\omega}$) is allowed to flow between two contacts (e.g. $1$ and $2$) of the monolayer branch outside the top gated region (Fig.~2a), setting up a temperature gradient ($\Delta T$) across the tBLG region. The resulting second-harmonic thermo-voltage ($V_{2\omega}$) is recorded between leads $3$ and $4$ as a function of doping and heating current (Fig.~2b)~\cite{zuev2009thermoelectric,mahapatra2017seebeck}. Different heating and measurement configurations yield similar variation of $V_{2\omega}$ with $n$, suggesting that the two layers are uniformly hybridized across the overlap area (see SI, section VII). The linear response was ensured from  $V_{2\omega} \propto I^{2}_{\omega}$ for the range of heating current used (Fig.~2d). As a function of $n$, $V_{2\omega}$ exhibits multiple sign-reversals as $E_\mr{F}$ is varied across the lowest energy bands which, at high temperatures ($T \approx 70$~K), align well with the derivative of $R$ expected from SMR (Eq.~\ref{Semiclassical Mott relation}) (Fig.~2c,d). While the sign reversals near CNP and the super-lattice gaps at $\nu = \pm 4$ are due to changes in the quasiparticle excitations, those near $\nu \approx \pm 2$ are attributed to the Lifshitz transitions due to the change of Fermi surface topology when the chemical potential is tuned across the vHS in the lowest energy band~\citep{cao2016superlattice,kim2016charge}. We speculate that the observed asymmetry in the zero-crossings of $V_{2\omega}$ at the Lifshitz transitions on the electron and hole sides is most likely related to the particle-hole asymmetry of the band structure itself (Fig.~1d).

The $n$-dependence of $V_{2\omega}$ deviates from that expected from SMR as $T$ is decreased below $\sim 40$~K. This is shown in Fig.~2e, where two new extrema, consisting of a maximum at $\nu = +2$ and minimum at $\nu = -2$, develop as $T$ is lowered. To compare with the SMR quantitatively, we rewrite Eq.~\ref{Semiclassical Mott relation} as,
\begin{equation}
\label{Mott relation}
S_\mr{Mott}=\frac{\pi^2 k_\mr{B}^2 T}{3|e|}\frac{1}{R}\frac{\mr{d}R}{\mr{d}V_\mr{tg}} \frac{\mr{d}V_\mr{tg}}{\mr{d}n} \frac{\mr{d}n}{\mr{d}E}\bigg\vert_{E_\mr{F}},
\end{equation}
where $(1/R)\mr{d}R/ \mr{d}V_\mr{tg}$ is measured experimentally, and $\mr{d}n/\mr{d}E$ is obtained from the calculated DOS in Fig.~1d ($\mr{d}V_\mr{tg}/\mr{d}n = e/C_\mr{hBN}$, where $C_\mr{hBN}$ is the known topgate capacitance per unit area). Using $\Delta T$ as the single fitting parameter, we obtain excellent agreement between the measured $V_{2\omega}$ and Eq.~\ref{Mott relation} at the CNP ($\nu = 0$) and $\nu = \pm4$ simultaneously which also confirms that $\Delta T$ is largely unaffected by doping of the tBLG region. This is a key advantage of our `crossed' device architecture that maintains heating efficiency by heating the ungated section of the same device~\citep{mahapatra2019misorientation}. While the SMR explains the observed $V_{2\omega}$ over almost the entire doping regime ($-4 \lesssim \nu \lesssim +4$) at high temperatures ($\gtrsim 40$~K) (bottom panel of Fig.~2e), the excess thermovoltage centered around $\nu = \pm2$, becomes evident at lower $T$. We also find evidence of small excess $V_{2\omega}$ between $\nu = -3$ and $-4$, but its comparison with SMR becomes inaccurate at high $T$ due to considerable thermal activation component in $R$ close to the super-lattice gap. Using the $\Delta T$ extracted from the fitting of $V_{2\omega}$, we show the $T$-dependence of $S = V_{2\omega}/\Delta T$ in Fig.~2f for different $n$ (see SI, sections VIII and IX). As is evident, $S$ exhibits a linear dependence on $T$ at all doping except in the vicinity of $\nu = \pm 2$. The $S\propto T$ behavior is expected in a degenerate weakly or non-interacting metal within the semiclassical framework, and has been verified for monolayer graphene~\citep{zuev2009thermoelectric} as well as tBLG at slightly larger $\theta$ ($2\degree \lesssim \theta \lesssim 5 \degree$)~\cite{mahapatra2019misorientation}. Close to $\nu =\pm 2$, $S$ exhibits a non-monotonic $T$-dependence that changes sign at $\approx 40$~K and, in contrast to the expectation of $S \approx 0$ from SMR, saturates to a non-zero magnitude $S \approx \pm2~\mu$V/K for $\nu=\pm2$ respectively, at low $T$ (inset of Fig.~2f). This is remarkable because, (1) at low $T$, the observed sign of $V_{2\omega}$ can not be assigned to the electron(hole)-like bands any more, and (2) the excess $S$ persists to a temperature scale ($\sim 40$~K) that is much higher than the superconducting transition ($T_\mr{c} \sim 1.7$~K) in tBLG at $\theta = \theta_\mr{c}$ or the temperature scale for correlated Mott-insulator ($\lesssim 4$~K)~\cite{cao2018correlated,cao2018unconventional,kerelsky2019maximized}, suggesting a very distinct nature of the ground state.

Although the Mott formula has been verified in a range of graphene-based devices \citep{zuev2009thermoelectric,jayaraman2020evidence}, it can be violated in the hydrodynamic regime \citep{ghahari2016enhanced} and due to phonon drag in cross-plane thermoelectric transport in tBLG at $\theta > 6^\circ$ \citep{mahapatra2017seebeck}. Nevertheless, these effects depend on the dominance of {\it e-e} and/or {\it e-}phonon scattering and hence appear only at higher temperatures ($> 100$~K). However, as shown in Fig.~3a, the occurrence of excess $S$, normalized as $(S-S_\mr{Mott})/S_\mr{max}$, where $S_\mr{max}$ is the maximum value of $S$ at a given $T$, is concentrated in dome-like regions around $\nu = \pm2$ in the $T-(\nu,n)$ phase diagram. It is known that $e-e$ interactions enhance the thermopower beyond the limit set by SMR \cite{behnia2015fundamentals}. For example, the enhanced $S$ in some correlated oxides \citep{wang2003spin}  has been attributed to spin entropy in many-body interacting states, while that in many of the heavy Fermions \citep{izawa2007thermoelectric} is attributed to shrinking of the Fermi surface close to quantum critical points where NFL effects dominate. A near-ubiquitous feature of the NFL regime in itinerant Fermionic systems, ranging from cuprates \cite{da2014ubiquitous}, ruthanates \cite{rost2009entropy}, pnictides \cite{lee2012non} to heavy Fermions \citep{izawa2007thermoelectric}, is the `strange metal' phase, characterized by the absence of well defined quasiparticles and linear $T$ dependence of $\rho$. Theoretical work also suggests possibilities of excess entropy, analogous to Bekenstein-Hawking entropy in charged black holes, in this regime, that remains finite down to vanishingly small $T$ \citep{sachdev2015bekenstein}. Furthermore, the $T$-linearity in $\rho$ corresponds to a scattering rate $\tau^{-1} \sim k_\mr{B}T/\hbar$, in the universal Planckian limit, as observed in many correlated oxides and heavy fermionic systems \citep{bruin2013similarity}, and recently claimed in tBLG at $\theta = \theta_\mr{c}$~\citep{cao2020strange}.

 To check the mutuality between the excess entropy and the strange metallic behaviour, we compare the $n$-dependence of normalized excess $S$ at $T = 5$~K (Fig.~3b), and the scattering rate obtained from the slope $\mr{d}\rho/ \mr{d}T$ in the $T$-dependence of $\rho$ (Fig.~3c). For reference, we also present the results from another device at $\theta \approx 4^\circ$, where we find no violation of SMR over the experimental range of $n$. In the NFL state, the incoherent scattering rate is $\tau^{-1} = Ck_\mr{B}T/\hbar$, where the dimensionless coefficient $C$ is of the order of unity for Planckian dissipation. In Fig.~3c we plot $n$-dependence of $\mr{d}\rho/ \mr{d}T$ and $C$, where $C$ is computed from $\mr{d}\rho/ \mr{d}T$ assuming Drude-like resistivity in accordance to Ref.~\citep{cao2020strange,bruin2013similarity} (See SI, section XI). Away from the CNP, $\mr{d}\rho/ \mr{d}T \approx 10$~$\Omega$/K is almost independent of $n$ upto $\nu \approx \pm4$, which is nearly two orders of magnitude larger than $\mr{d}\rho/ \mr{d}T \approx 0.2 - 0.3$~$\Omega$/K  for the tBLG device at $\theta \sim 4^{\circ}$, implying that the individual layers are essentially decoupled in the latter~\citep{cao2020strange,polshyn2019large}. Intriguingly, for tBLG at $\theta = 1.6^\circ$, we find $C$ to approach the order of unity in the vicinity of $\nu \to \pm 2$, raising the possibility of a common physical origin as the violation of SMR. We have restricted  the calculation of $C$ upto $\nu =\pm 2$ to avoid artefacts originating from the effective doping ($n_\mr{c}$) used in calculating $C$, which is not proportional to the filling factor everywhere in the phase-diagram \citep{cao2020strange}.

To understand the origin of excess $S$ theoretically, we explored the impact of electron interaction and vHS within a dynamical mean field theory (DMFT)~\cite{Georges1996,Yuan2019,Haldar2018}. Considering the four lowest bands near the CNP and a Hubbard interaction $U=0.2~W$, we find that the low-energy vHSs enhance the effect of interaction for fillings $|\nu|\simeq 1-2$, with a low coherence temperature scale, below which the system behaves as FL (see schematic of Fig.~4a, Methods and SI, sections XII, XIII for details). The strong self-energy effects near the vHSs lead to deviations of $S$ from the non-interacting or high-temperature thermopower around $|\nu|\simeq 1-2$ (Fig.~4b), as well as sign changes as a function of $T$ (Fig.~4c), that are qualitatively similar to the experimental observations. However, the DMFT results seems unable to capture the apparent saturation to finite $S (\nu=\pm2)$ at low $T$ (Fig.~2f, inset) as well as the persistence of $T$-linear $\rho$ down to the lowest $T$ ($\approx 100$~mK) at $\nu = \pm 2$ in the experiment (Fig.~1g). 

Since both theoretical \citep{gonzalez2017electrically} and experimental \citep{Lin2019Magnetism} investigations claim magnetic textures in low-angle tBLG near $\nu = \pm2$, we measured the thermoelectric response in the presence of a large in-plane magnetic field. Fig.~4e shows no appreciable change in the thermovoltage $V_{2\omega}$ measured at $3$~K for in-plane magnetic fields upto $9$~T. Thus we conclude that, unlike the superconducting and Mott insulating states \citep{cao2018correlated,cao2018unconventional,lu2019superconductors}, the violation of SMR is not sensitive to underlying spin degeneracy. The nonmagnetic excess $S$ may arise from the correlation-induced $U(1)$ valley symmetry breaking, and the scattering of electrons with the Goldstone modes in the inter-valley coherent (IVC) ordered state~\cite{Po2018}. While such an effect may persist till higher $T$ ($\sim 40$~K), the scattering with Goldstone modes is not expected to give rise to strong violation of SMR as in the case of usual electron-phonon scattering~\cite{Jonson1990}. Nevertheless, the lifting of valley degeneracy, provides an estimate of $A = \mr{d}\rho/ \mr{d}T \sim h/2e^2W \approx 6.2$~$\Omega$/K that closely matches the experimental observation (Fig.~3c), providing likely evidence of interaction-dominated transport \cite{cao2020strange}.

In summary, we have measured the electrical resistivity and thermopower in twisted bilayer graphene for twist angle $\theta \approx 1.6^{\circ}$ at various temperatures. Our experimental results show concurrent $T$-linear resistivity at Planckian dissipation scales and emergent thermopower below $T\lesssim 40$~K at near $\nu = \pm 2$ that results in the breakdown of semiclassical Mott relation. The thermopower near $\nu = \pm 2$ approaches a finite magnitude ($\approx 2$~$\mu$V/K) at low $T$ providing a new facet to the strongly correlated `strange metal' phase in tBLG. Our experimental results point to a truly non-Fermi liquid (NFL) metallic state in tBLG at low twist angle that carry strong similarities to those observed in cuprates or heavy-Fermion materials with low coherence temperatures.

The authors thank Nano mission, DST for the financial support. M.J. and S.M. thank the computational facilities in SERC. K.W. and T.T. acknowledge support from the Elemental Strategy Initiative conducted by the MEXT, Japan, Grant Number JPMXP0112101001,  JSPS
KAKENHI Grant Numbers JP20H00354 and the CREST(JPMJCR15F3), JST. U.C. acknowledges funding from IISc and SERB (ECR/2017/001566), and H.R.K from SERB(SB/DF/005/2017). S.B. acknowledges funding from IISc and SERB (ECR/2018/001742). 

B.G. and P.S.M. contributed equally to this work.

\section{Methods}

\subsection{ Device fabrication}

All devices in this work were fabricated using a layer-by-layer mechanical transfer method \citep{mahapatra2017seebeck}. Monolayer graphene and hexagonal boron nitride (hBN)  were exfoliated on SiO$_{2}$/Si wafers and graphene edges were identified using optical microscopy and Raman spectroscopy. The edges of the graphene flakes were aligned under an optical microscope and encapsulated within two hBN layers to prevent the channel from disorder and to act as dielectric for electrostatic gating. Electron beam lithography was used to define Cr/Au top gate for tuning the number density in tBLG region. Finally, the electrical contacts were patterned by electron-beam lithography and reactive ion etching followed by metal deposition (5 nm Cr/50 nm Au) using thermal evaporation technique.

Electrical transport measurements were performed in a four-terminal geometry  with typical \textit{ac} current excitations of $10$-$100$~nA using a standard low-frequency lock-in amplifier at $226$~Hz, in a dilution refrigerator and a $1.5$-K cryostat. For thermoelectric measurements, local Joule heating was employed to create a $\Delta T$ across the tBLG channel. A range of sinusoidal currents ($2$-$5$~$\mu$A) at excitation frequency $\omega$~$=$ $17$~Hz were used for Joule heating and the resulting $2^\mr{nd}$ harmonic thermal voltage ($V_{2\omega}$) was recorded using a lock-in amplifier. Thermoelectric measurements were conducted in $1.5$-K cryostat with magnetic field of upto $9$~T. 

\subsection{Tight binding calculation of DOS}
	 The rigid bilayer structures were generated using the Twister code \cite{naik2018ultraflatbands}. The structures were subsequently relaxed in LAMMPS \cite{lammps}\cite{lammpsurl} using REBO \cite{rebo} as the intralayer potential and DRIP \cite{drip} as the interlayer potential. These relaxed structures were used for performing all the calculations. \newline
	 The electronic band structures were calculated by approximating the tight binding transfer integrals under the Slater Koster formalism \cite{slaterkoster}. A more detailed discussion on the calculations is available in the SI, section V.
	 
\subsection{DMFT calculations}
For the calculations of thermopower in DMFT, we assume a description of the  four bands near the CNP in terms of an effective low-energy hexagonal lattice model on the lattice \cite{Koshino2018,Po2018,Po2019}. Each hexagonal lattice site has two electronic orbitals and two spins ($\sigma=\pm 1/2$) indexed by $\alpha=1,\dots,4$, such that there are four bands that can hold a maximum of eight electrons per triangular unit cell of the hexagonal lattice. We further assume a $SU(4)$ symmetric on-site repulsive Hubbard interaction, namely 
\begin{align}
\mathcal{H}=-\sum_{ij,\alpha} t_{ij} c_{i\alpha}^\dagger c_{j\alpha}+U\sum_{i,\alpha<\gamma} n_{i\alpha}n_{i\gamma}\label{eq:LatticeHubbardModel}
\end{align} 
Here $c_{i\alpha}$ is the electron operator for $i$-th hexagonal lattice site and $n_{i\alpha}=c_{i\alpha}^\dagger c_{i\alpha}$. The hopping integrals are in general complex and can be chosen to fit \citep{Koshino2018,Po2018,Kang2018} the energy dispersion from band-structure calculation, e.g. as shown in Fig.~1d of the main text.  Within the DMFT approximation, discussed in detail in the SI, only the DOS of the low-energy bands enter and we take the DOS directly from our full tight-binding band-structure calculation discussed in the main text. The justification of using the above lattice model and estimations of the interaction strength is given in the SI. In the DMFT, the above lattice model is reduced to an effective single-site Anderson impurity hybridized with a bath whose properties are self-consistently determined using the non-interacting lattice DOS \citep{Georges1996} and the local impurity Green's function. We use a modified multi-orbital iterative perturbation theory (IPT) \citep{Kajueter1996,Dasari2016} impurity solver which has been benchmarked \cite{Dasari2016} previously with numerically exact continuous-time quantum Monte Carlo solver \citep{Gull2011}. Once the electronic self-energy is known from the DMFT, the thermopower is calculated using the standard formula \citep{Palsson1998}. The latter requires the transport DOS as an input, which is obtained from the energy dispersion of the four low-energy bands near the CNP. The details of the calculations are discussed in the SI, sections XII and XIII.

\bibliographystyle{naturemag}
\bibliography{Ref}

 \end{document}


\title{Supplementary information: Excess entropy and breakdown of semiclassical description of thermoelectricity in twisted bilayer graphene close to half filling}

\author{Bhaskar Ghawri}
\email{gbhaskar@iisc.ac.in}
\author{Phanibhusan S. Mahapatra}
\email{phanis@iisc.ac.in}
\author{Shinjan Mandal}
\author{Aditya Jayaraman}

\affiliation{Department of Physics, Indian Institute of Science, Bangalore, 560012, India}
\author{Manjari Garg}
\affiliation{Department of Instrumentation and Applied Physics, Indian Institute of Science, Bangalore, 560012, India}
\author{K. Watanabe}
\author{ T. Taniguchi}
\affiliation{National Institute for Materials Science, Namiki 1-1, Tsukuba, Ibaraki 305-0044, Japan}
\author{H. R. Krishnamurthy}

\author{Manish Jain}
\author{Sumilan Banerjee}
\affiliation{Department of Physics, Indian Institute of Science, Bangalore, 560012, India}
\author{U. Chandni}
\affiliation{Department of instrumentation and Applied Physics, Indian Institute of Science, Bangalore, 560012, India}
\author{Arindam Ghosh}
\email{arindam@iisc.ac.in}
\affiliation{Department of Physics, Indian Institute of Science, Bangalore, 560012, India}
\affiliation{Centre for Nano Science and Engineering, Indian Institute of Science, Bangalore 560 012, India}

\pacs{}
\maketitle

\subsection{\fontsize{12}{15}\selectfont I. Device characterization}

Fig. S1(a) shows the optical micrograph of a typical twisted bilayer graphene (tBLG) device fabricated for this study. We have used Raman spectroscopy to qualitatively differentiate between large twist angle and small twist angle tBLG devices. Specifically, the shape and position of the $2D$ peak is sensitive to the twist angle and provides a rough estimation\citep{havener2012angle,kim2012raman,mahapatra2017seebeck}. Fig. S1(b) shows the Raman spectra of $\sim 1.6\degree$ , $\sim 2.4\degree$, $\approx4\degree$ tBLG devices and single layer graphene (SLG). $2D$ peak of small twist angle devices have higher values of the full width half maximum (FWHM) than SLG and usually have an additional shoulder. In contrast, large twist angle tBLG has a spectrum very similar to SLG as the two layers are decoupled. Though, the Raman spectrum does not have enough resolution to precisely determine the twist angle, it provides a qualitative information about the angle and $\theta \sim 1.6\degree$ , $\theta \sim 2.4\degree$ are estimated based on transport data, while $\theta \approx4\degree$ is a rough estimate based on the shape of the $2D$ peak observed in Raman spectra.

\begin{figure}[H]
  \includegraphics[width=1.0\textwidth]{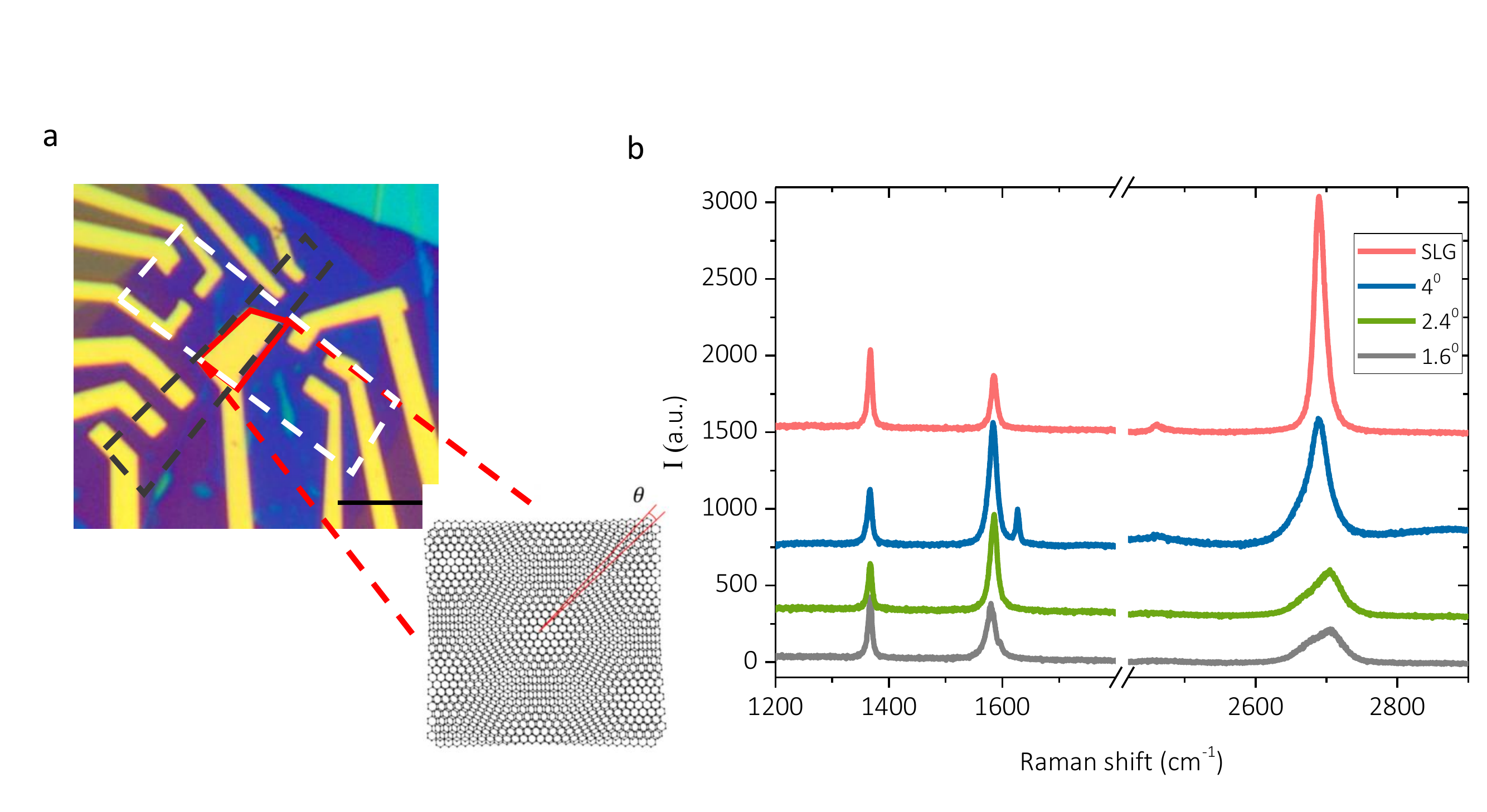}
  \captionsetup{justification=raggedright,singlelinecheck=false}
\caption{(a) Optical micrograph of a typical tBLG device. The scale bar represents a length of $5$~$\mu$m and dotted lines show the graphene edges. (b) Comparison of Raman spectra for G peak and 2D peak obtained for $\approx 1.6\degree$ , $\sim 2.4\degree$, $\approx4\degree$ and single layer graphene with relative offset in the intensity for clarity. }
\end{figure}

\subsection{\fontsize{12}{15}\selectfont II. Resistance measured  in different configurations for $1.6\degree$ device}

We have measured the electrical resistance in multiple contact configurations to probe the twist angle inhomogeneity in our device. Optical micrograph of the device along with the contact configuration is shown in Fig. S2(a), where the dashed lines show the two graphene layers and the metal top gate. Fig. S2(b) shows a comparison of four terminal resistance measured at $3$~K in three different configurations. Even though the magnitude of the resistance is slightly different in various configurations , the device exhibits a consistent moir\'{e} unit size in all configurations, as is evident by the good alignment of insulating states ($\nu= \pm4$), thereby ensuring angle homogeneity. Further, we have measured the temperature dependence of resistance  in the configuration given by, $I: 3-8, V:2-9$. Fig. S2(c),(d) show the resistance in the $T$ linear regime. We find that $R$ is metallic at $T \lesssim T_\mr{H}$, where $T_\mr{H} \sim 100 - 200$~K is a doping-dependent characteristic temperature. The results obtained in this configuration are consistent with the transport data shown in Fig. 1. Furthermore, it is to be noted that results shown in Fig. S2(c),(d) are from a different thermal cycle as compared to Fig. S2(b). 

\begin{figure}[H]
 \centering
  \includegraphics[width=1.0\textwidth]{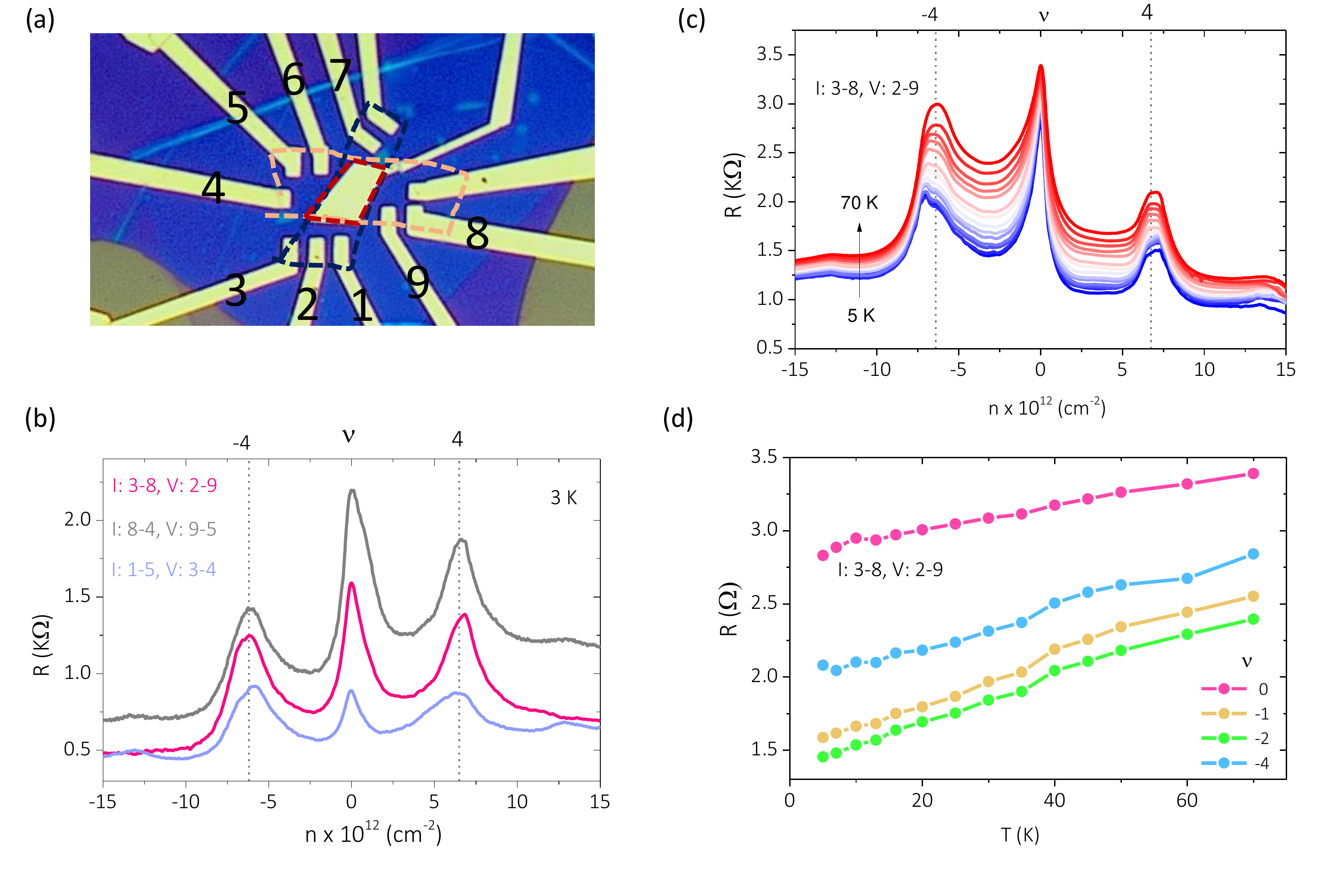}
 \captionsetup{justification=raggedright,singlelinecheck=false}
\caption{(a) Optical micrograph of the device with contact configuration (b) Measured resistance ($R$) for three different configurations. Dashed lines show filling factor $\nu = \pm 4$. (c) $R$ as a function of density at different $T$ ranging from $5$~K to $70$~K in the configuration marked by, $I: 3-8, V:2-9$. (d) T-dependence of $R$ at different filling factors ($\nu$) in the $T$-linear regime.}
\end{figure}

\subsection{\fontsize{12}{15}\selectfont III. Electrical transport in different thermal cycles}

Fig. S3(a) shows the transport data from the $1.6\degree$ device in one of the thermal cycles (labelled as `a'), where the resistance shows a shallow peak near $\nu=\pm2$. We observe a clear electron-hole asymmetry in the $R$ peaks, which is likely to emerge from the asymmetric band structure of tBLG. Similar shallow peaks in resistance reported in previous studies are attributed to van Hove singularities in the non-interacting DOS which in-turn lead to large scattering cross-section for quasiparticles \citep{kim2016charge,chung2018transport}. However, we notice that the $R$ peaks are suppressed by the application of a parallel magnetic field and finally vanishes at a field of $6(2)$~Tesla for $\nu= 2(-2)$ (Fig. S3(b)), which is similar to the effects seen in previous studies on magic angle tBLG \citep{cao2018correlated}. Furthermore, we have measured the T-dependence in a small range to probe the origin of these peaks (Fig. S3(c)). Although the magnitude of the peak decreases at higher $T$ ($7-8$~K), it is not possible to extract any quantitative information from the $T$ dependence because of limited data. However, these features almost vanish after a thermal cycle to the sample (labelled by `b') as shown in Fig. S3(d). Observation of  such features in $R$ along with a clear violation of Mott formula indicate the role of interactions, however the origin of these peaks in $R$ is still not very clear at this point.

\begin{figure}[H]
\centering
  \includegraphics[scale=0.6]{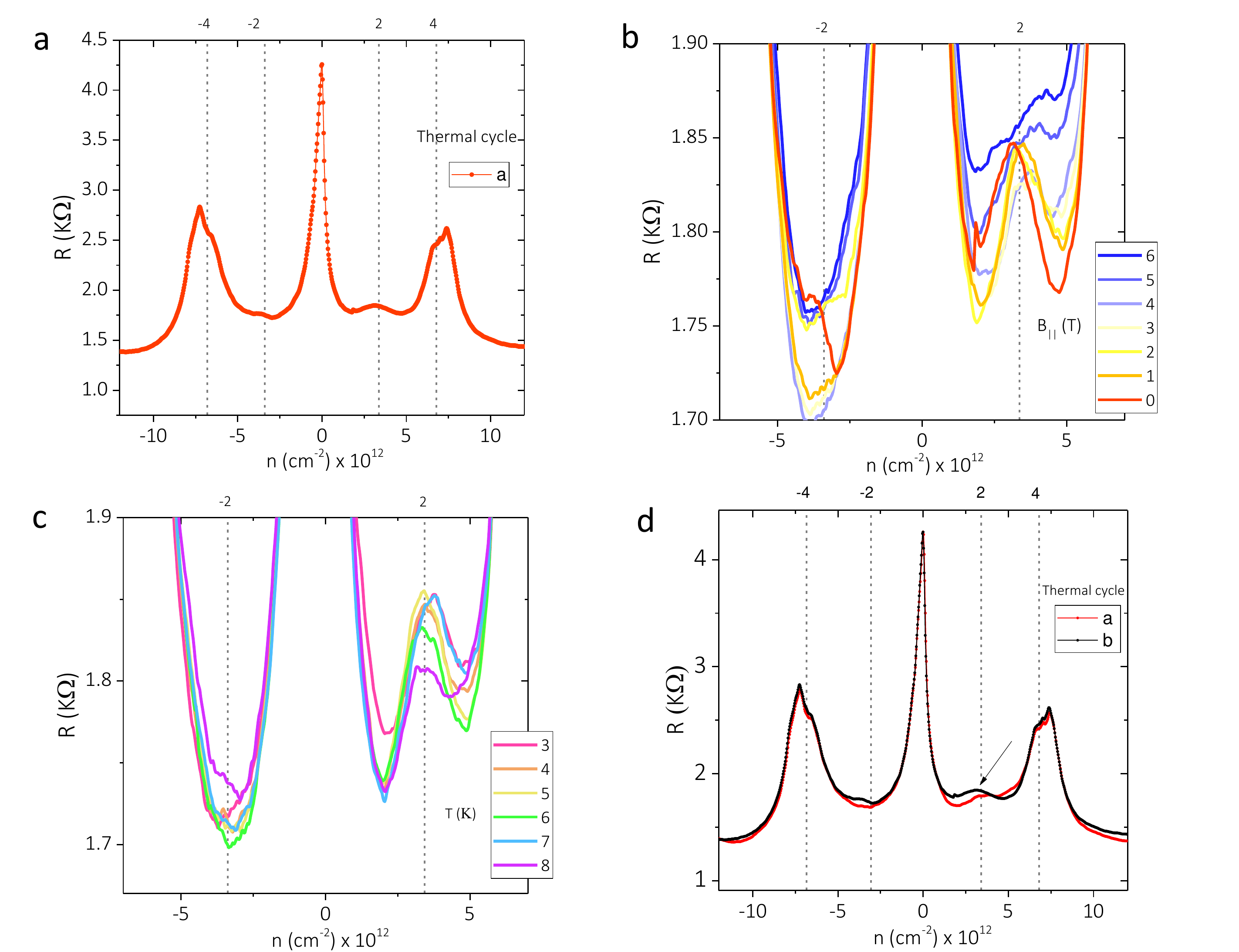}
 \captionsetup{justification=raggedright,singlelinecheck=false}
\caption{. (a) $R$ as a function of of $n$ for in thermal cycle number `a' (b) Parallel Magnetic field dependence of $R$, zoomed-in to focus on $R$ peak at $\nu=\pm 2$. (c) $T$ dependence of resistance peak near $\nu=\pm2$ (d) $R$ plotted for two different thermal cycles.  }

\end{figure}
\subsection{\fontsize{12}{15}\selectfont IV. Quantum oscillations in $1.6\degree$ device}
Fig. S4(a) shows the resistance ($R$) measured at $5$~K in the absence of an external magnetic field, and Fig. S4(b) shows the results of quantum oscillations measurements in $1.6\degree$ device . We have plotted the first derivative of $R$ with respect to the gate voltage tuned from Dirac point ($V_{tg}-V_{D}$) in order to get a better colour contrast. The quantum oscillations emerging from the CNP are eight fold degenerate as previously reported for devices with similar twist angles\citep{cao2016superlattice}. In contrast, the Landau fans emerging from $\nu=\pm4$ are four fold degenerate, accounting for spin and Fermi-contour degeneracy. Additionally, we have estimated the twist angle using the quantum oscillations, which comes out to be $\approx 1.6\degree$ 

\begin{figure}[H]
  \includegraphics[width=1.0\textwidth]{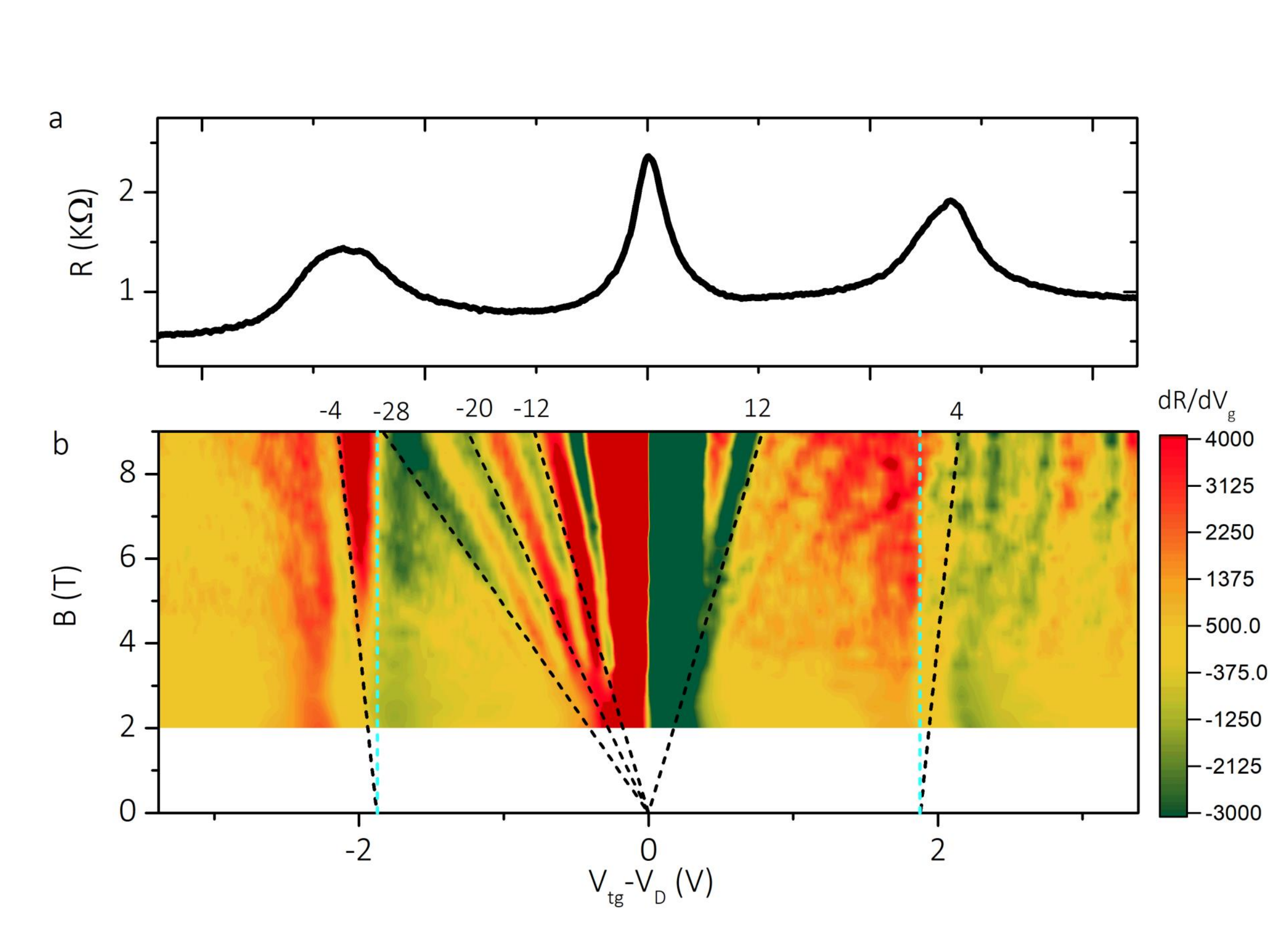}
  \captionsetup{justification=raggedright,singlelinecheck=false}
\caption{ (a) $R$ as a function of $n$ measured at $5$~K. (b) Quantum oscillation in $1.6\degree$ device. The first derivative of $R$  with respect to gate voltage is plotted in order to enhance the colour contrast. Black dashed lines show the Landau fans emerging from the CNP and $\nu=\pm4$.  } 
\end{figure}

\subsection{\fontsize{12}{15}\selectfont V. Computational Details for the tight binding formalism}
	
\begin{figure}[H]
	\centering
	\includegraphics[scale=0.3]{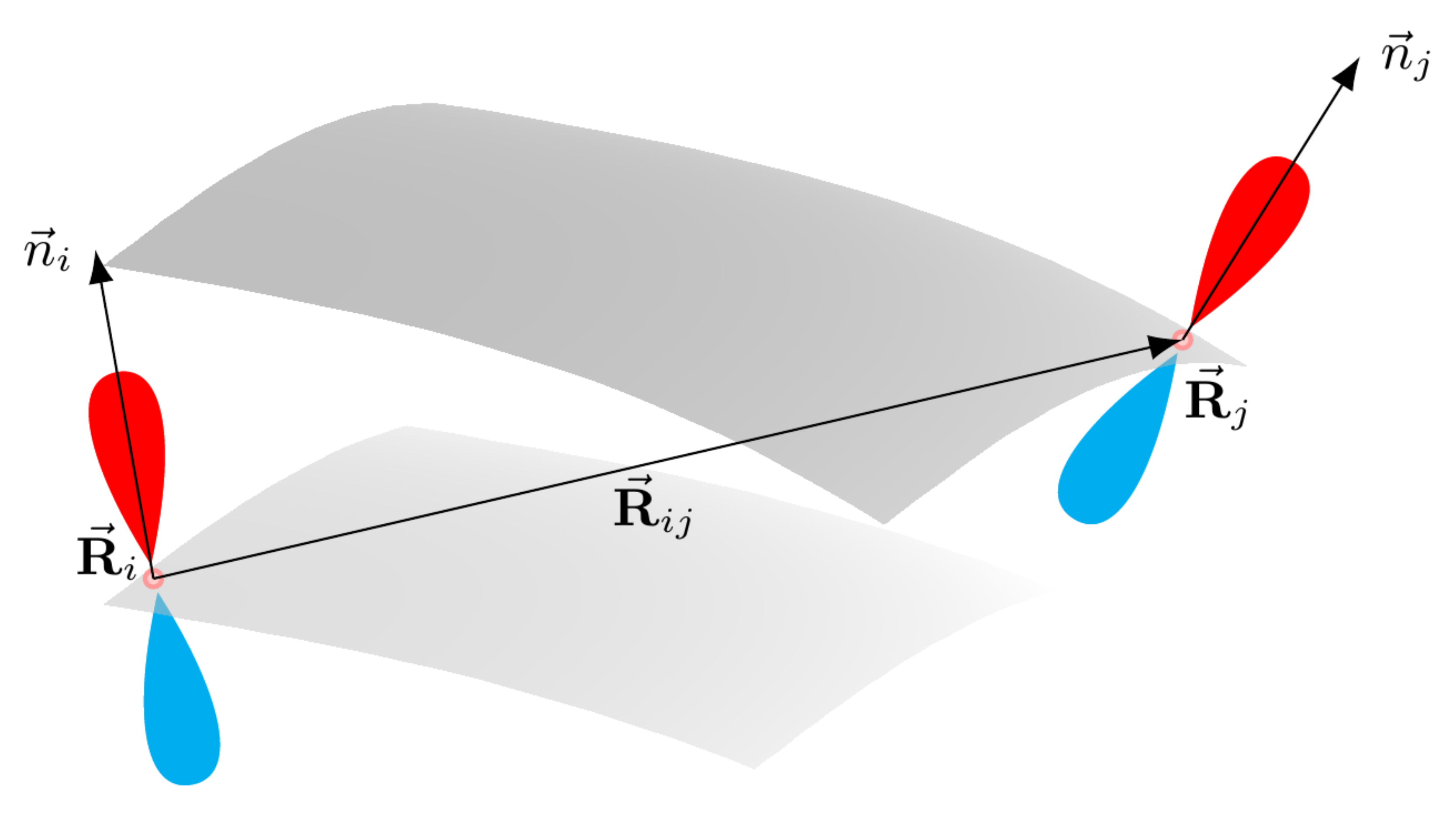}
	\caption{Local normals at $\mathbf{R}_i$ and $\mathbf{R}_j$. Note that, $\hat{\mathbf{n}}_{i/j} = \frac{\vec{\mathbf{n}}_{i/j}}{| \vec{\mathbf{n}}_{i/j}|}$ and $\hat{\mathbf{R}}_{ij} =\frac{\vec{\mathbf{R}}_{ij}}{| \vec{\mathbf{R}}_{j}|}$ }
	\label{locnorm}
\end{figure}

	The electronic hamiltonian of the system is written as 
	\begin{equation}
	\hat{\mathbf{H}} = -\sum_{i,j} t(\mathbf{R}_i-\mathbf{R}_j)c_i^{\dagger}c_j + \text{h.c.} = -\sum_{i,j} t_{ij}c_i^{\dagger}c_j + \text{h.c.}	
	\end{equation}
	where $\mathbf{R}_i$ denotes the real space position of the $i^\text{th}$ atom, and $c_i^{\dagger}$ and $c_i$ are the creation and annihilation operators at $\mathbf{R}_i$. We approximate the transfer integrals $t_{ij}$ under the Slater Koster formalism \cite{slaterkoster} assuming that the overlap of the $p_z$ orbitals can be approximated as the linear combination of the $\sigma\sigma$ and the $\pi\pi$ overlaps. 
	Taking the local curvature of the sheets into account the transfer integral can be written as \cite{skchoicnt}:
	\begin{align}
	t_{ij}  =& t_{\pi\pi}[\hat{\mathbf{n}}_i-(\hat{\mathbf{n}}_i\cdot\hat{\mathbf{R}}_{ij})\hat{\mathbf{R}}_{ij}]\cdot[\hat{\mathbf{n}}_j-(\hat{\mathbf{n}}_j\cdot\hat{\mathbf{R}}_{ij})\hat{\mathbf{R}}_{ij}]+t_{\sigma\sigma}[\hat{\mathbf{n}}_i\cdot\hat{\mathbf{R}}_{ij}]\cdot[\hat{\mathbf{n}}_j.\hat{\mathbf{R}}_{ij}]
	\label{eqn2}
	\end{align}
	where $\hat{\mathbf{n}}_i$ is the unit normal at the $i^{\text{th}}$ site, and $\hat{\mathbf{R}}_{ij}$ is the unit vector joining the sites.\newline
	We can see that from fig.(\ref{locnorm}) if there is no local curvature, and: \vspace{-0.05in}\begin{itemize}
			\item[i.] if $i$ and $j$ are on the same layer then the only non-zero contribution is from the $\pi\pi$ term
			\item[ii.] if i and j are on different layers then $t_{ij} = t_{\sigma\sigma}\cos^2(\theta)+t_{\pi\pi}\sin^2(\theta)$ where $\theta$ is the angle that $\hat{\mathbf{n}}_i$ makes with $\hat{\mathbf{R}}_{ij}$  
	\end{itemize}
	The terms $t_{\pi\pi}$ and $t_{\sigma\sigma}$ are taken as follows:
	\begin{equation}
		\begin{aligned}
		&t_{\pi\pi} = t_{\pi}^0\exp\Big(-\frac{|\vec{\mathbf{R}}_{ij}| - a_0}{\delta}\Big); \ \ \ \ t_{\sigma\sigma} = t_{\sigma}^0\exp\Big(-\frac{|\vec{\mathbf{R}}_{ij}| - d_0}{\delta}\Big)
		\end{aligned}
	\end{equation}
	On setting the parameter $a_0 = 1.42 \ \text{\AA}$, the nearest neighbour distance between two Carbon atoms, it is reasonable to take $t_{\pi}^0 \approx -2.7 \ \text{eV}$, the nearest neighbour transfer energy in monolayer graphene. Similarly by taking $d_0 = 3.35 \ \text{\AA}$, the interlayer distance in AA stacked bilayer, we fix $ t_{\sigma}^0 \approx \ 0.48 \ \text{eV}$. 
	The attenuating factor, $\delta = 0.184a_0$ is chosen such that the strength of the next nearest neighbour is $0.1$ times the nearest neighbour interaction in monolayer graphene.\cite{moon2012energy}.\newline
	To compute the density of states we use an analog of the linear tetrahedron method for 2D systems, the linear triangulation method. A uniform $(110 \times 110)$ grid is taken in the Brillouin Zone (BZ) and a delaunay triangulation is performed on those points. The density of states and the number density are then estimated by integrating within each of the triangles.
	
	\subsection{\fontsize{12}{15}\selectfont VI. Additional transport data for the $1.6\degree$ device }

Fig. S6(a),(b) show $\rho$ in the $T$-linear regime across a wide range of carrier density between the CNP and $\mp n_{s}$. In this $T$ range, transport can be assumed to be restricted to the lowest electron-and hole- superlattice subbands. The dashed lines are linear fits to $\rho(T)$ . A quantitative analysis of the fitted slope versus the density is shown in Fig. 3c. Furthermore, an insulating behaviour near $\pm n_s$ is observed in electrical transport as $T$ is increased above $\approx 90$~K. To study the thermally activated transport behaviour of insulating states, we plot temperature dependence of resistance at $\nu=\pm4$ in Fig. S7(a),(b). An Arrhenius-like behaviour is clearly evident in this temperature range. From the slope in the Arrhenius plot, we estimate the activation gaps to be $\sim 160$ and $\sim 240$~K for electron-side and hole-side respectively.

 \begin{figure}[H]
 \centering
  \includegraphics[scale=0.55]{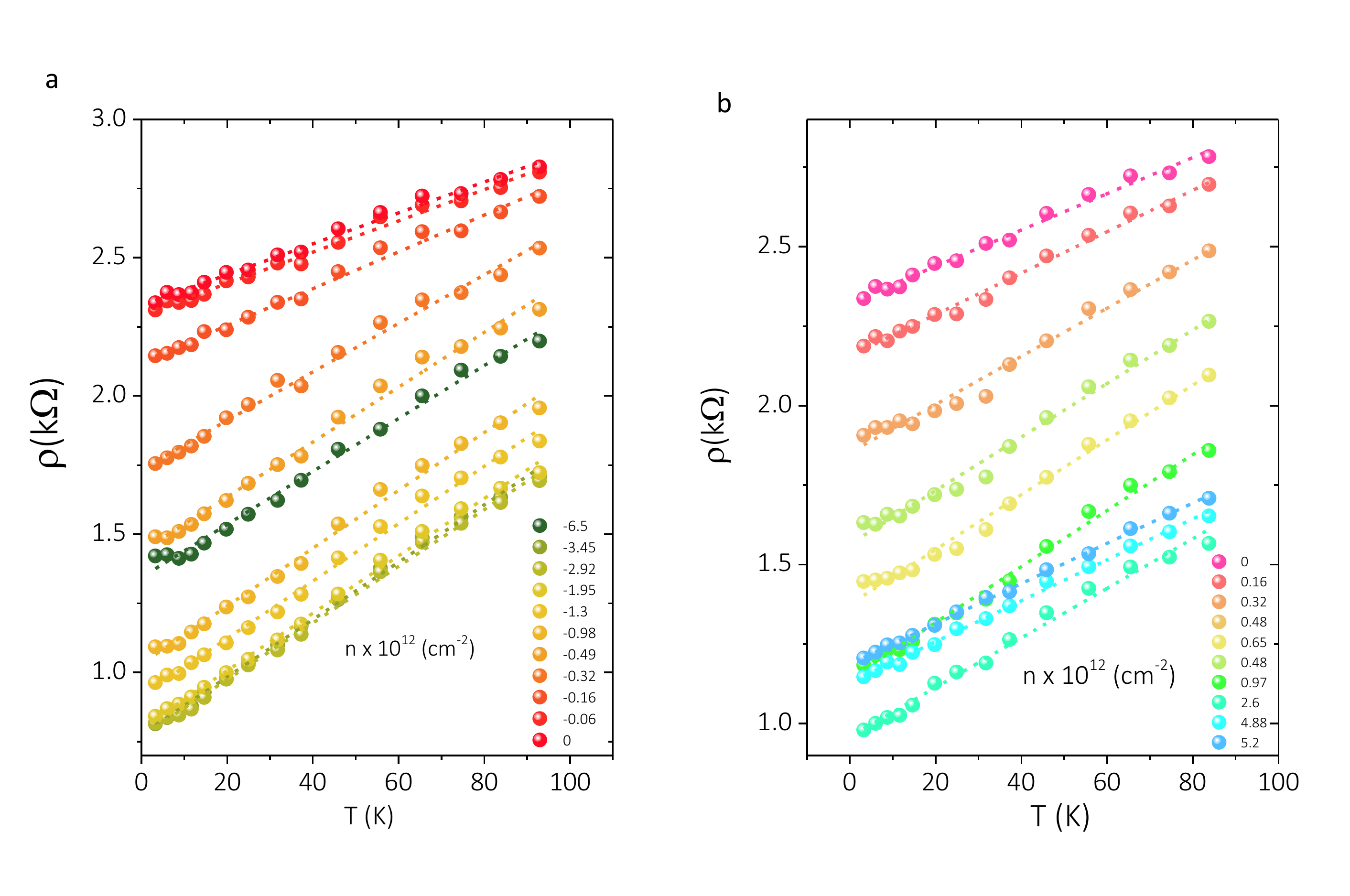}
 \captionsetup{justification=raggedright,singlelinecheck=false}
\caption{(a),(b) $\rho$ as a function of $T$ for selected values of $n$ in T-linear regime. Dashed lines show linear fit to the data.}
\end{figure}
 
 \begin{figure}[H]
 \centering
  \includegraphics[scale=0.55]{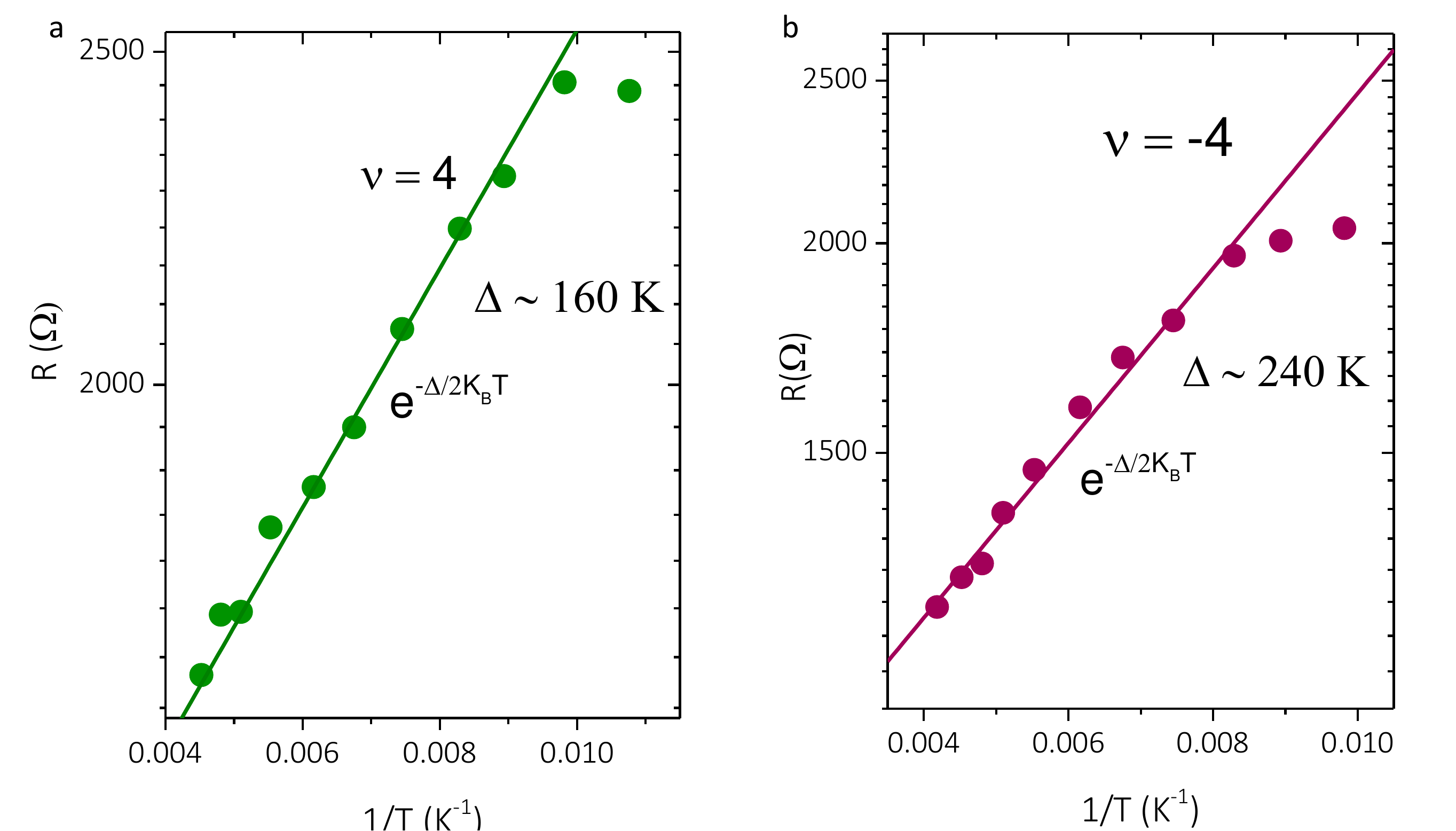}
 \captionsetup{justification=raggedright,singlelinecheck=false}
\caption{(a),(b) $R$ as a function of $T$ plotted in T-linear regime. Dashed lines show linear fit to the data.}
\end{figure}

\subsection{\fontsize{12}{15}\selectfont VII. Thermoelectric (TE) measurements in different heating configurations in $1.6\degree$ device }

We have performed thermoelectric measurements in different heating configurations to verify the results described in Fig. 2. The optical micrograph of the device along with the contact configuration is shown in Fig. S8(a). Briefly, a sinusoidal current is passed between two contacts to create a $\Delta T$ in the tBlG region and the resulting second harmonic voltage $V_{2\omega}$ is measured. Fig. S8(b) shows the TE measured in three different configurations at $5$~K. We note that apart from the sign reversal at the CNP and the vHS, two new extrema, consisting of a maximum at $\nu = +2$ and minimum at $\nu = -2$ develop, which is very similar to the results described in main text (Fig. 3c). However, a slight asymmetry in different configurations is observed, which can arise from local inhomogeneity in the twist angle.  Furthermore, we have performed temperature dependent TE measurements in the configuration given by, $I: 1-9, V:3-4$. Fig. S9 shows $V_{2\omega}$ for six different temperatures. We note that as $T$ is increased above $35-40$~K, the excess TE  near $\nu=\pm2$ vanishes, indicating the $T$ scale upto which correlation effects are present.  Reproducibility of our TE results in different heating configurations ascertains the distinct nature of the ground state

\begin{figure}[H]
 \centering
  \includegraphics[scale=0.7]{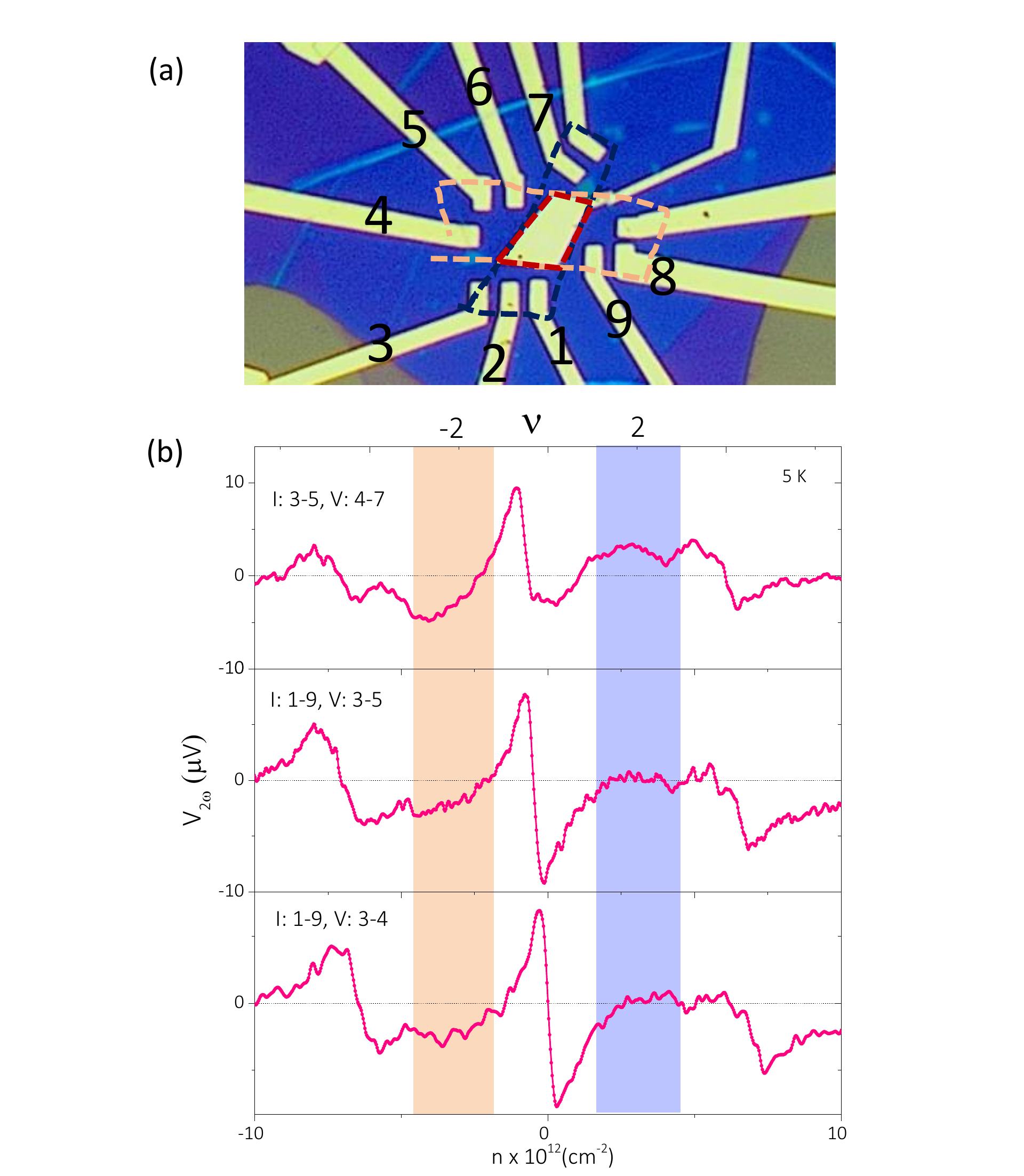}
 \captionsetup{justification=raggedright,singlelinecheck=false}
\caption{(a) Optical micrograph of the device with contact configuration (b) Measured $V_{2\omega}$ for three different heating configurations.}
\end{figure}

\begin{figure}[H]
  \includegraphics[width=1.0\textwidth]{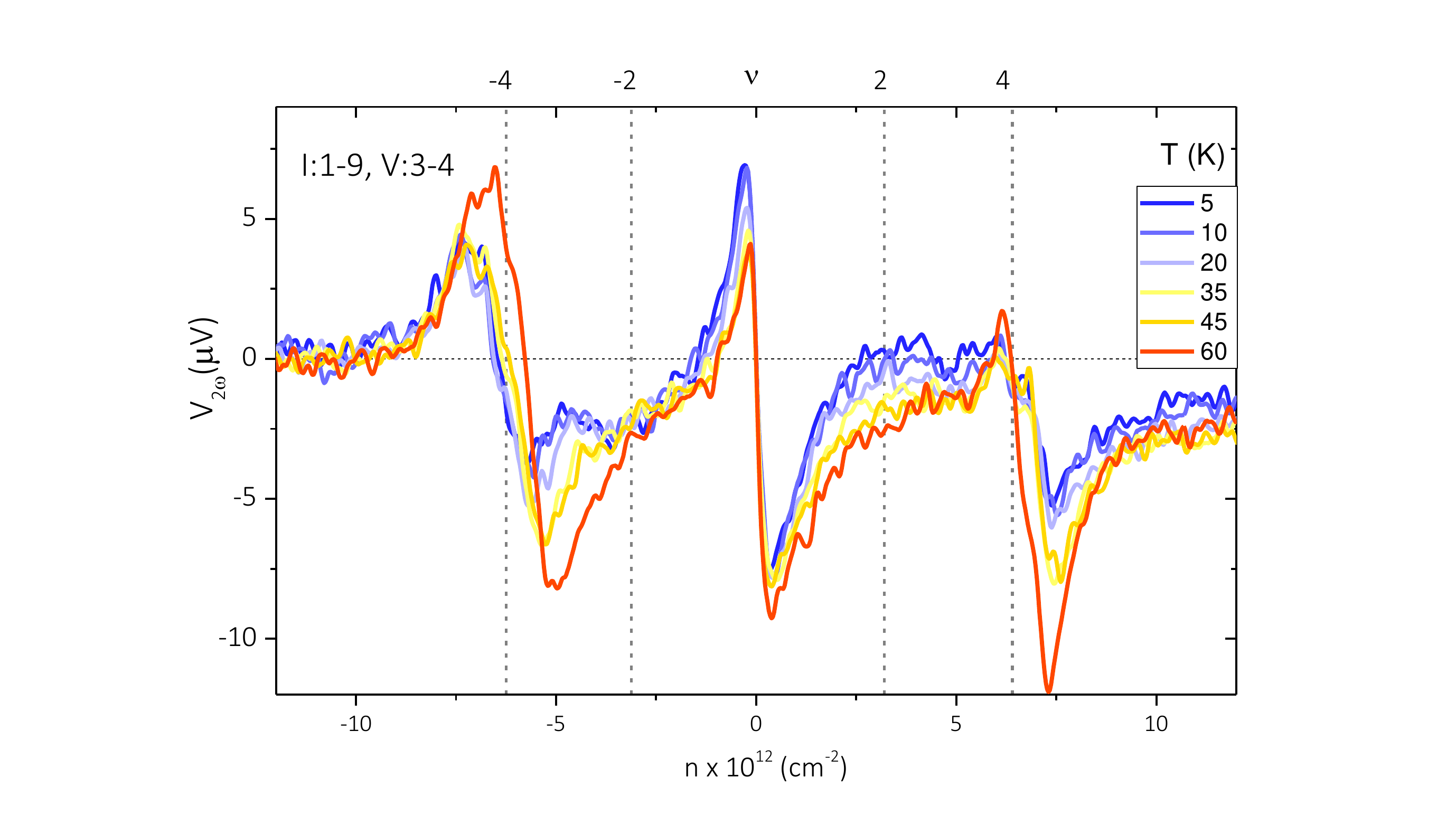}
 \captionsetup{justification=raggedright,singlelinecheck=false}
\caption{ Measured $V_{2\omega}$ for six representative $T$ in configuration, $I: 1-9, V:3-4$ . Dashed lines show filling factors.}
\end{figure}

\subsection{\fontsize{12}{15}\selectfont VIII.  Seebeck coefficient fitted with Mott formula  for $1.6\degree$ device}
The fitting of the Mott formula ($S_\mr{Mott}$) obtained using the calculated density of states (DOS) for three different twist angles with the measured thermopower for the $1.6\degree$ device is depicted in Fig. S10. We observe that $S_\mr{Mott}$ from the DOS for $1.61\degree$ matches well with measured $V_{2\omega}$ both at the CNP and $\nu=\pm4$ simultaneously, whereas if the twist angle is tuned away from $1.6\degree$, we find a deviation in Mott fit at $\nu=\pm4$ as clearly shown for $1.69\degree$ and $1.41\degree$. This further emphasises that the thermoelectric transport is extremely sensitive to the band structure and in turn any correlation effects, which can not be probed with conductance measurements alone.
\begin{figure}[H]

  \includegraphics[width=1.0\textwidth]{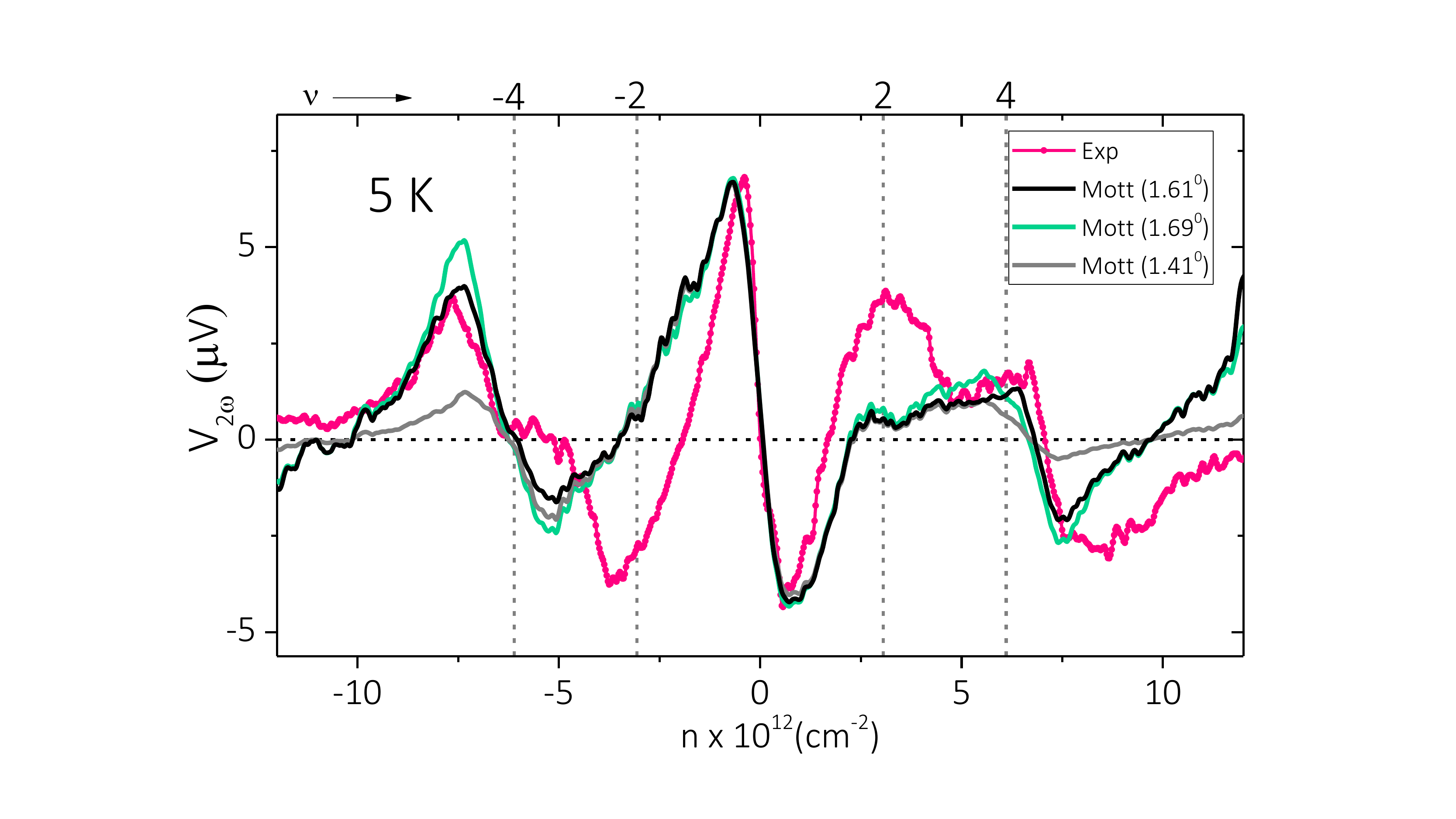}
  \captionsetup{justification=raggedright,singlelinecheck=false}
\caption{ Doping dependence of measured $V_{2\omega}$ compared with $S_\mr{Mott}$ calculated using DOS for three different twist angles $1.61\degree$,$1.69\degree$ and $1.41\degree$. }
\end{figure}
\subsection{\fontsize{12}{15}\selectfont IX. $\Delta T$ calibration for the $1.6\degree$ device}
The thermoelectric power (TEP) or Seebeck coefficient $S$ is obtained from the ratio of thermoelectric voltage and the temperature gradient $(V_{2\omega}/\Delta T)$ across the tBLG. In our earlier work on tBLG \citep{mahapatra2017seebeck,mahapatra2019misorientation}, we have used graphene resistance thermometry for obtaining $\Delta T$, however, at lower $T$, this method is not very reliable because of a weak dependence of graphene resistance on the temperature . Hence we employ Mott formula to estimate $\Delta T$. This method of obtaining $\Delta T$ relies on the assumption that the system follows the Mott formalism, which is indeed  true for low angle tBLG \citep{mahapatra2019misorientation}. Furthermore, the Mott formula exhibits an excellent match with the experimental data both at the CNP and $\nu = \pm4$ simultaneously (Fig. 2d), which in turn justifies our method. We have fit the Mott formula to $V_{2\omega}$ at different values of $T$ and obtained $\Delta T$ as shown in Fig. S11. 

\begin{figure}[H]
 \centering
  \includegraphics[scale=0.8]{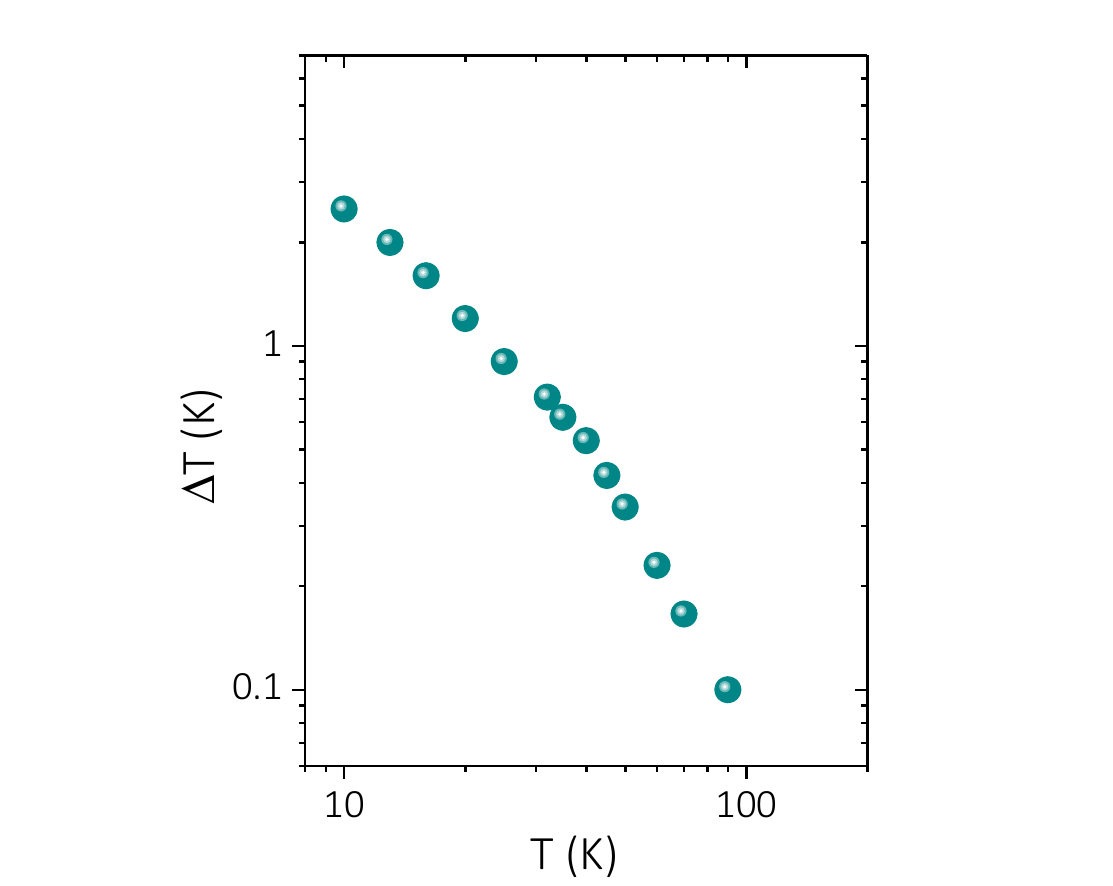}
\caption{ The extracted values of $\Delta T$ obtained from Mott fit at various $T$.}
\end{figure}
\subsection{\fontsize{12}{15}\selectfont X. Cross-plane resistance vs. $T$ and comparison of $S$ with calculated $S_\mr{Mott}$for $\theta \approx 4 \degree$ }
Fig. S12(a) shows the T-dependence of cross-plane resistance ($R_{CP}$) for various $n$ for $\theta\sim 4\degree$. Except near the Dirac point where electron-hole puddles dominate, it shows a weak metallic T-dependence with $d\rho/ dT$ $\approx 0.1$~$\Omega$/K, which is notably very different from $\theta \sim 1.6 \degree$ device, where we observe metallic behaviour at all $n$. Furthermore, Fig. S12(b) shows a comparison of the density dependence of the measured $S$ with $S_\mr{Mott}$ evaluated using a single layer Dirac dispersion. $S_\mr{Mott}$ shows an excellent fit to the data, further suggesting that the two graphene layers are essentially decoupled at low energies.

\begin{figure}[H]
  \includegraphics[width=\columnwidth]{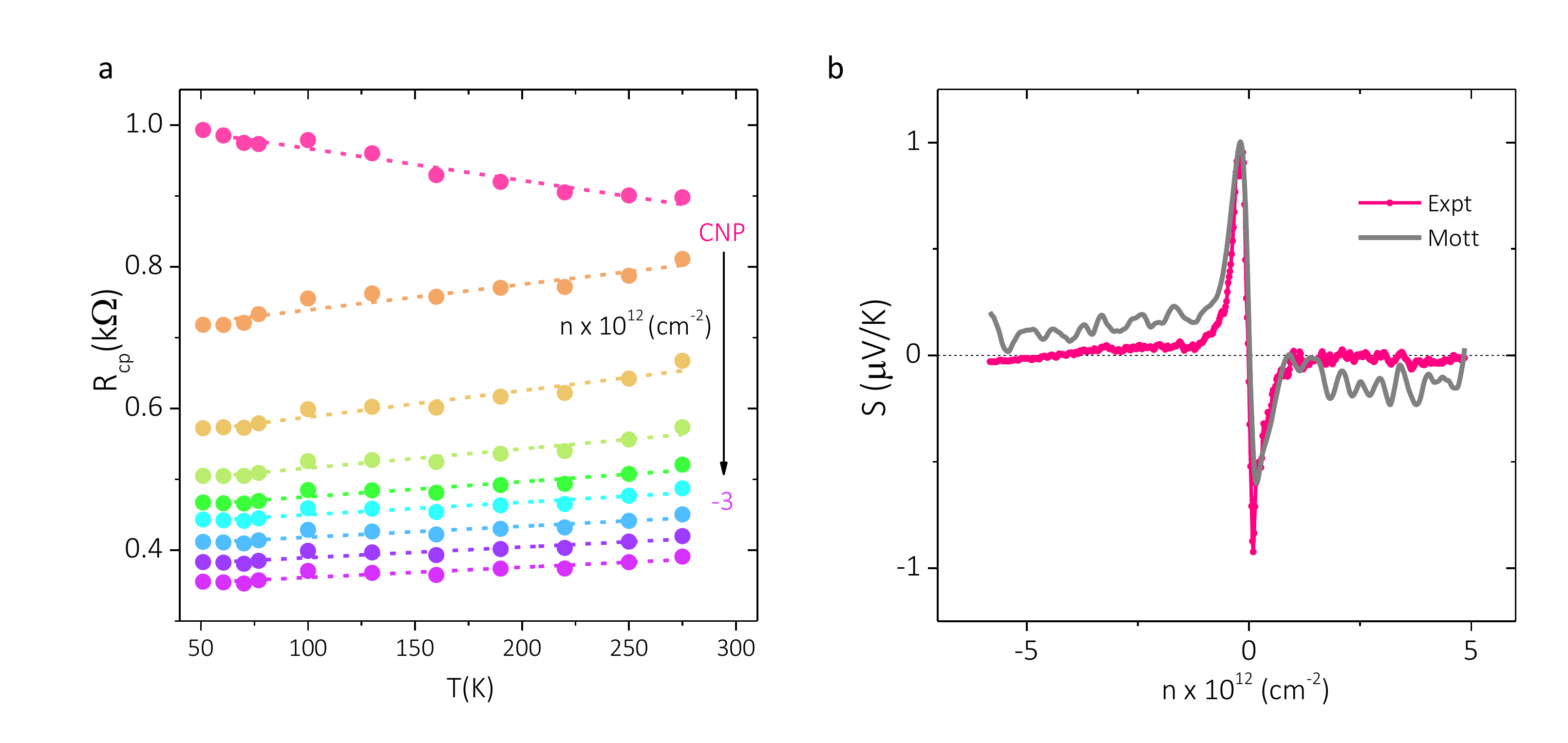}
  \captionsetup{justification=raggedright,singlelinecheck=false}
\caption{\textbf{ Thermoelectric transport for $\theta \approx 4 \degree$}.(a) Cross-plane resistance $R_{cp}$ as a function of $T$ for various $n$. The dashed lines show linear fits to the data. (b) The doping dependence of the measured $S$ (Pink circles) compared with $S_\mr{Mott}$ (gray solid lines) at $5$~K}.
\end{figure}
\subsection{\fontsize{12}{15}\selectfont XI. Planckian dissipation}
In non-Fermi Liquid state (NFL), the incoherent scattering is often found to be bound by the Planckian dissipation scale $\tau^{-1} = Ck_\mr{B}T/\hbar$, where the dimensionless coefficient $C$ is order of unity.
\begin{equation}
\label{equation 2}
C=\frac{\hbar}{k_\mr{B}}\frac{e^{2}  n_\mr{c}}{m^{*}(0)} A
\end{equation}

where $n_\mr{c}$ is the gate-induced carrier density and $m^{*}(0)$ is cyclotron mass at $T \rightarrow 0$ which is proportional to the density of states per Fermi pocket at the Fermi energy for two dimensional materials. We calculate the effective mass ($m^{*}$) as a function of $n$ in the devices with twist angles $1.6\degree$ and $4\degree$ using the expression, $m^{*}= (h^{2}/2\pi) D(E)/N$, where $N$ is the degeneracy. We have used N=8 (as evident from the Landau fan) and  density of states calculated by tight binding method for $1.6\degree$ and the SLG DOS for $4\degree$. Fig. S13(a),(b) show the calculated  $m^{*}$ for the two devices. We note that our calculated values of $m^{*}$ for $1.6\degree$ matches well with previously measured value for a device with similar twist angle \citep{polshyn2019large}. The above expression allows us to calculate $C$ and hence the scattering rate $\tau^{-1}$ over a range of number density as shown in Fig.~3c (right axis). We find that $C\sim$~$\mathcal{O}$~$(1)$ for $\theta \approx 1.6\degree$ near $\nu = \pm 2$, where the deviation from Mott relation is maximum. In comparison, the value of $C$ is two orders of magnitude smaller for $\theta \sim 4^{\circ}$.

\begin{figure}[H]
  \includegraphics[width=1.0\textwidth]{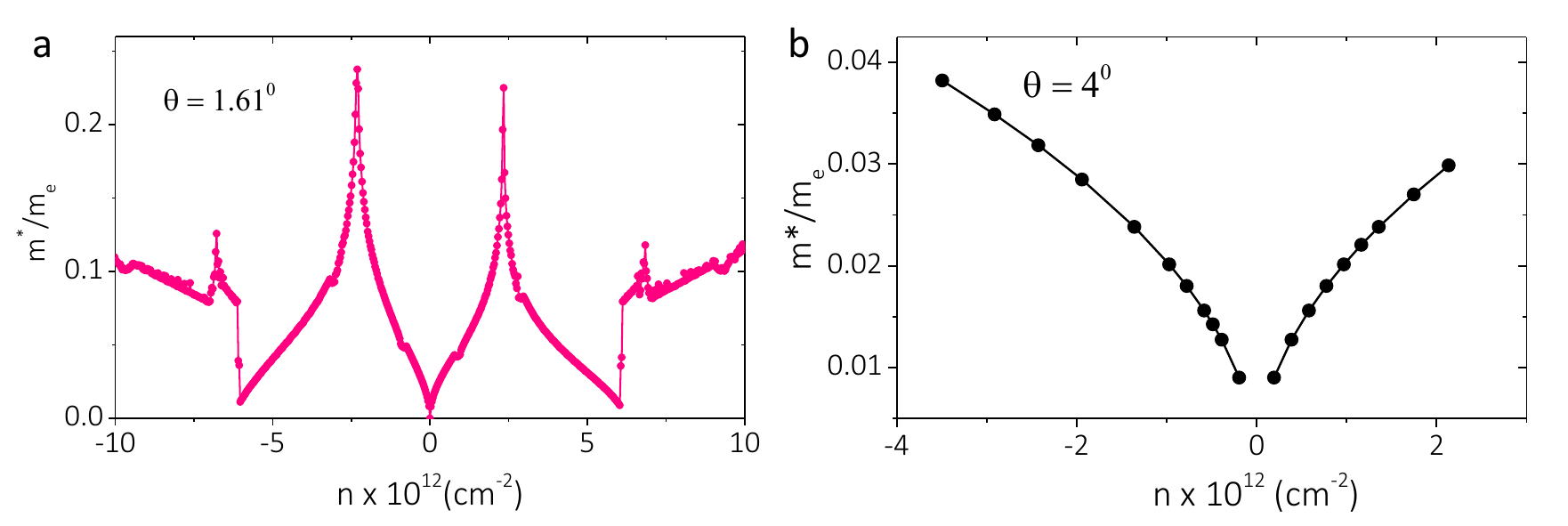}
  
\caption{Numerically calculated cyclotron mass for (a) $\theta = 1.61 \degree$ and (b) $\theta \sim 4 \degree$}. 
\end{figure}

\subsection{\fontsize{12}{15}\selectfont XII. Lattice model for the low-energy bands} 
To set up the dynamical mean field theory (DMFT) calculations involving the four low-energy bands near the CNP we assume an effective hexagonal lattice model for the moire lattice with two electronic orbitals and two spins ($\sigma=\pm 1/2$) indexed by $\alpha=1,\dots,\mathcal{M}$ with $\mathcal{M}=4$, such that there are four bands that can hold a maximum of eight electrons per triangular unit cell, i.e.
\begin{align}
\mathcal{H}=-\sum_{ij,\alpha} t_{ij} c_{i\alpha}^\dagger c_{j\alpha}+\mathcal{H}_{int} \label{eq:LatticeModel}
\end{align} 
Here $c_{i\alpha}$ is the electron operator for $i$-th hexagonal lattice site. The hopping integrals are in general complex and can be chosen to fit \cite{Koshino2018,Po2018,Kang2018} the energy dispersion from band-structure calculation, e.g. that shown in Fig.1d of the main text. However, the two-orbital model is known \cite{Po2018,Po2019} to be insufficient to reproduce the topology of the Bloch wave functions while keeping all the low-energy symmetries of the continuum model \cite{Bistritzer2011}. Within the simplified DMFT approximation discussed below only the DOS of the low-energy bands enter and we take the DOS directly from our full tight-binding bandstructure calculation discussed in the main text. Hence the tight-binding parameterization for low-energy effective lattice model does not explicitly appear in our DMFT calculations. We also neglect the asymmetry between orbitals induced by the  band structure. The interaction term $\mathcal{H}_{int}=U\sum_{\hexagon}(\sum_{i\in\hexagon,\alpha}n_{i\alpha}/3-4)^2$ is the cluster Hubbard term \cite{Koshino2018,Po2018} where the sum runs over all the hexagons in the moire lattice and $n_{i\alpha}$ is the electron number operator for $\alpha$ orbital/spin on the $i$-th site. The cluster term arises since the physical charges are concentrated around the AA regions of the moire lattice. When expanded the cluster Hubbard term leads to on-site as well as longer-range interactions with strength comparable to the on-site one \cite{Koshino2018}. The longer-range terms could be important for various possible symmetry-breaking orders \cite{Koshino2018,Po2018,Xu2018,Yuan2019} at low temperature near magic angles. Finding the phase diagram of the cluster Hubbard model for arbitrary filling is an interesting theoretical problem \cite{Xu2018,Yuan2019}. 

However, for the metallic state considered here within DMFT, the extended terms are mainly expected to lead to screening and renormalization of the on-site Hubbard $U$ \cite{Chitra2000}. To this end, we take 
\begin{align}
\mathcal{H}_{int}=U\sum_{i,\alpha<\gamma}n_{i\alpha}n_{i\gamma}, \label{eq:Hubbard}
\end{align}
i.e. an $SU(4)$-symmetric on-site Hubbard interaction. The minimal model already captures the crucial effects of interaction near the low-energy VHS, as we discuss here and in the main text. The strength of the on-site $U$ could be estimated as $U=1.857(e^2/4\pi\varepsilon\epsilon_0L_\mathrm{M})$ \cite{Cao2018a,Koshino2018}, where the moire lattice constant $L_\mathrm{M}=a_0/(2\sin(\theta/2))=8.89$ nm for graphene lattice spacing $a_0=0.246$ nm and twist angle $\theta=1.6^\circ$.
Hence $U\simeq 15-75$ meV depending on the dielectric constant $\varepsilon\simeq 20-4$ or screening due to the gates. Based on our estimated total band width of $W=180$ meV for the four low-energy bands near CNP, we get a moderate interaction strength $U/W\simeq 0.1-0.4$. For our DMFT calculations, we take $U/W\simeq 0.2$. We find that the effect of this moderate interaction gets enhanced near the low-energy van Hove singularities (VHS).

\subsection{\fontsize{12}{15}\selectfont XIII. Dynamical mean field theory}
In the DMFT approximation \cite{Georges1996}, and assuming a homogeneous state with orbital and spin ($SU(4)$) symmetry, we reduce the model of Eq.\eqref{eq:LatticeModel} to an effective single-site impurity problem with the imaginary-time action 
\begin{align}
S_{imp}&=-\int_0^\beta d\tau d\tau'\sum_\alpha \mathcal{G}^{-1}(\tau-\tau')\bar{c}_\alpha(\tau)c_\alpha(\tau')\nonumber \\
&+U\sum_{\alpha<\gamma}\int_0^\beta d\tau n_\alpha(\tau)n_\gamma(\tau) \label{eq:ImpurityAction}
\end{align}
Here $\beta=1/k_\mathrm{B}T$ and $(\bar{c}_\alpha,c_\alpha)$ are fermionic Grassmann variables with $n_\alpha=\bar{c}_\alpha c_\alpha$. The dynamical mean field $\mathcal{G}^{-1}(\ci\omega_n)=\ci\omega_n+\mu-\Delta(\ci \omega_n)$ for the Matsubara frequency $\omega_n=(2n+1)k_\mathrm{B}T$, with $n$ being an integer, is determined by the hybridization function $\Delta(\ci\omega_n)$ which is self-consistently determined using the non-interacting lattice DOS as we discuss below. The chemical potential is fixed by the filling.

Since we need the real-frequency electronic Green's function $G(\omega)=G(\ci\omega_n\to\omega+i0^+)$ to compute dc transport coefficients, e.g. Seebeck coefficient, we use an approximate impurity solver, namely modified iterative perturbation theory (IPT) \cite{Georges1996,Kajueter1996} and its generalization for the multi-orbital case \cite{Dasari2016}. The latter has been benchmarked with numerically exact techniques for solving the single-impurity problem \cite{Dasari2016} and is expected to work quite well for moderate interaction strengths, like in our case. Within the IPT, we obtain the impurity self-energy as
\begin{align}
\Sigma(\omega)&=(\mathcal{M}-1)U\langle n\rangle+\frac{A(\mathcal{M}-1)\Sigma^{(2)}(\omega)}{1-B(\mathcal{M}-1)\Sigma^{(2)}(\omega)}, \label{eq:IPT_SelfEnergy}
\end{align}
and the impurity Green's function is obtained from the Dyson equation $G^{-1}(\omega)=\mathcal{G}^{-1}(\omega)-\Sigma(\omega)$. The first term in Eq.\eqref{eq:IPT_SelfEnergy} is the Hartree self energy and $\Sigma^{(2)}(\tau)=U^2\tilde{\mathcal{G}}^3(\tau)$ is the second-order self-energy obtained using the Hartree-corrected impurity Green's function $\tilde{\mathcal{G}}^{-1}(\omega)=\mathcal{G}^{-1}(\omega)-U\langle n\rangle$. The coefficients $A$ and $B$ are chosen to satisfy certain sum rules and the known high-frequency behaviour of the impurity Green's function \cite{Dasari2016} and are given by
\begin{align}
A&=\frac{\langle n\rangle (1-\langle n\rangle)}{\langle n_0\rangle (1-\langle n_0\rangle)}+\frac{(\mathcal{M}-2)(\langle nn\rangle -\langle n\rangle^2))}{\langle n_0\rangle (1-\langle n_0\rangle)}\\
B&=\frac{(1-(\mathcal{M}-1)U\langle n\rangle)+\mu_0-\mu}{(\mathcal{M}-1)U^2\langle n_0\rangle (1-\langle n_0\rangle)}
\end{align}
\begin{figure}[H]
	\centering
	\includegraphics[width=\linewidth]{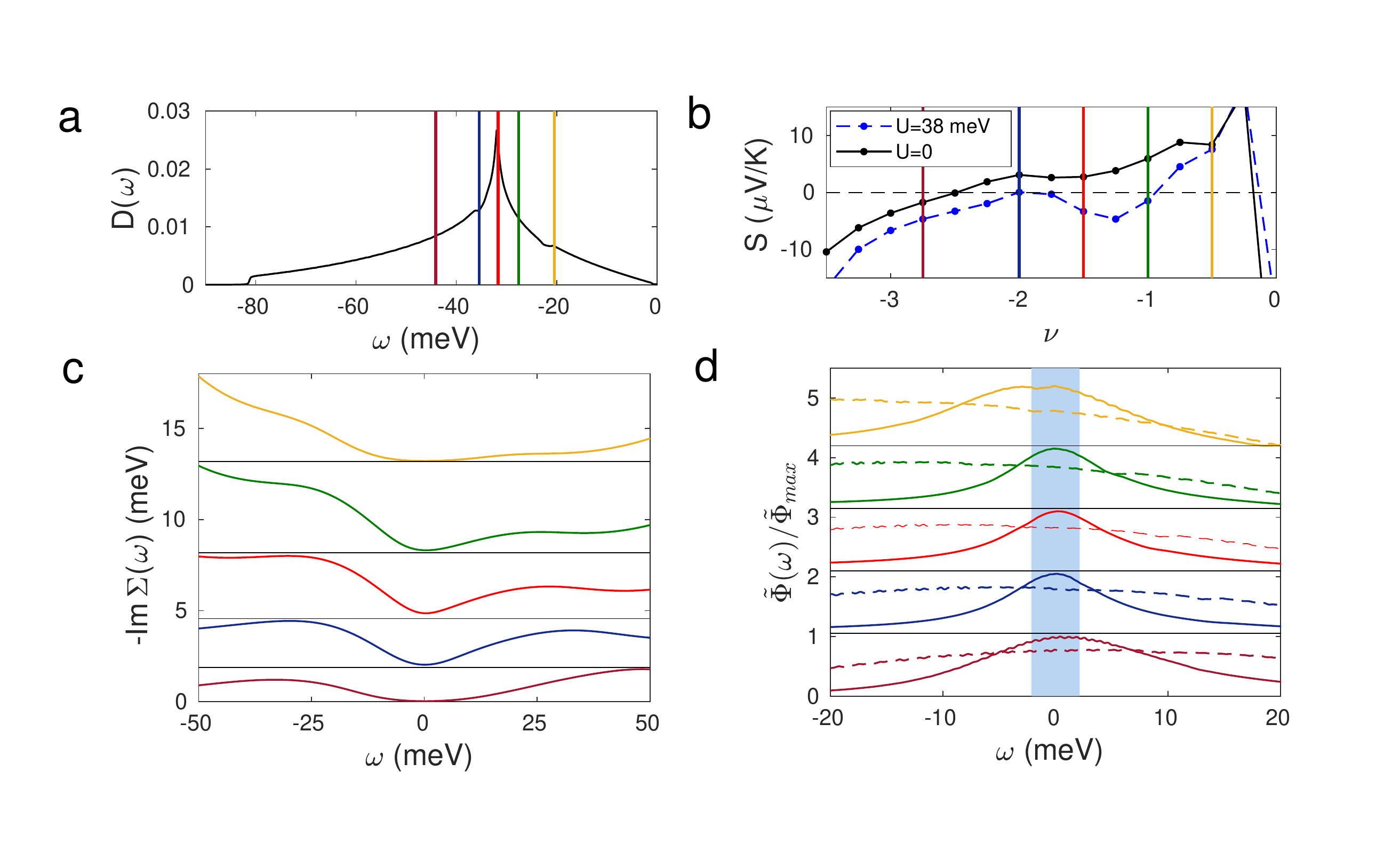}
	\captionsetup{justification=raggedright,singlelinecheck=false}
	\caption{{\bf Van Hove singularity, self-energy and the sign of thermopower :} (a) Non-interacting DOS from the tight binding bandstructure calculation for densities (or chemical potentials) below the CNP for $T=14$ K. The vertical lines indicate five densities across the low-energy van Hove singularity (VHS). (b) The same densities or fillings ($\nu$) are indicated by the vertical lines for the Seebeck coefficient ($S$) vs. $\nu$ plot at $T=14 K$ for the both non-interacting ($U=0$, black line with filled circles) and the interacting ($U=38$ meV, blue lines with filled circles) cases. The sign changes for the interacting and non-interacting cases happen at different fillings. (c) Imaginary part of electronic self energy at $T=14$ K for the five fillings shown in (a) and (b) as indicated by the colors. For clarity, the self-energies at the subsequent fillings are vertically shifted with the baselines indicated by the solid horizontal lines. The electronic self-energy is enhanced near the VHS with marked low-energy particle-hole asymmetry. The Fermi liquid-like behavior, $-\mathrm{Im}\Sigma(\omega)\sim \omega^2$, at low energies persist over a narrower energy window near the VHS indicating a more correlated metallic state with a lower coherence temperature scale. (d) Effective transport DOS $\tilde{\Phi}(\omega)$ (solid lines) for the same five fillings as in (a), (b) and (c), indicated by the same colors. Again, $\tilde{\Phi}(\omega)$  for the five subsequent fillings are vertically shifted for clarity. The large self-energy effects near the VHS strongly modify the effective transport DOS for the interacting case compared to the non-interacting $\tilde{\Phi}(\omega)$ (dashed lines). The shaded region has a width $\sim k_\mathrm{B}T$ determined by $-T(\partial n_\mathrm{F}(\omega)/\partial \omega)$ that controls the integral [Eq.\eqref{eq:Seebeck_A}] appearing in Eq.\eqref{eq:Seebeck} for the Seebeck coefficient. The interacting $\tilde{\Phi}(\omega)$ becomes sharply peaked near the VHS. The sign of $S$ depends on whether $\tilde{\Phi}(\omega)$ increases or decreases with $\omega$ near $\omega=0$ (see text for a discussion).} 
	\label{fig:SelfEnergy_S}
\end{figure}
Here $\langle n\rangle=-(1/\pi)\int_{-\infty}^\infty d\omega n_\mathrm{F}(\omega)\mathrm{Im}G(\omega)=(\nu+4)/8$ is the occupancy of a single orbital. The density-density correlation function $\langle nn\rangle\approx-\int_{-\infty}^\infty d\omega n_\mathrm{F}(\omega)\mathrm{Im}(\Sigma G)/(\pi U (\mathcal{M}-1))$ with $n_\mathrm{F}(\omega)=1/(e^{\beta\omega}+1)$, the Fermi function. Following ref.\onlinecite{Potthoff1997}, we fix the `pseudo' chemical potential $\mu_0$ from $\langle n_0\rangle=-(1/\pi)\int_{-\infty}^\infty d\omega n_\mathrm{F}(\omega)\mathrm{Im}\tilde{\mathcal{G}}(\omega)=\langle n\rangle$ using the Hartree corrected Green function $\tilde{\mathcal{G}}^{-1}(\omega)=\omega+\mu_0-\Delta(\omega)-U\langle n\rangle$.

Once the self-energy [Eq.\eqref{eq:IPT_SelfEnergy}] is obtained, the DMFT self-consistency condition is used to compute the local lattice Green's function
\begin{align}
G(\omega)&=\frac{1}{2\mathcal{M}}\int_{-\infty}^\infty d\epsilon\frac{D(\epsilon)}{\omega+\mu-\epsilon-\Sigma(\omega)},
\end{align}
assumed to be the same as the impurity Green's function for a single orbital and spin species. Here $D(\epsilon)$ is the total tight-binding DOS per triangular moire unit cell for the four lowest bands. The self-consistency loop is closed by obtaining the new hybridization function $\Delta(\omega)=\omega+\mu-\epsilon-\Sigma(\omega)-G^{-1}(\omega)$. The loop is iterated till we reach convergence. 

We fix the interaction strength $U=38$ meV and perform the calculations to obtain electronic Green's function and self-energy for fillings $-3<\nu<3$ and temperatures $5 \mathrm{K}\lesssim T\lesssim 100$ K. The Seebeck coefficient \cite{Palsson1998} $S$ is given by,
\begin{align}
S&=-\frac{k_\mathrm{B}}{e}\frac{A_1}{A_0} \label{eq:Seebeck}
\end{align}
where
\begin{align}
A_m&=\int_{-\infty}^\infty d\omega d\epsilon \rho^2(\epsilon,\omega)\Phi(\epsilon)\left(-T\frac{\partial n_\mathrm{F}(\omega)}{\partial \omega}\right)(\beta\omega)^m \label{eq:Seebeck_A}
\end{align}
with $m=1,2$. $\rho(\epsilon,\omega)=-(1/\pi)\mathrm{Im}(1/(\omega+\mu-\epsilon-\Sigma(\omega)))$ is obtained from DMFT self-energy and $\Phi(\epsilon)=(1/\mathcal{A})\sum_{n=1,\mathbf{k}}^4(\partial \epsilon_{n\mathbf{k}}/\partial k_x)^2\delta(\epsilon-\epsilon_{n\mathbf{k}})$ is the transport DOS that is obtained from tight-binding energy dispersion $\epsilon_{n\mathbf{k}}$ for the four bands near CNP. Here $\mathcal{A}$ is the area of the sample. To analyze the sign of the Seebeck coefficient, we define an effective transport DOS, $\tilde{\Phi}(\omega)=\int_{-\infty}^\infty d\epsilon \rho^2(\epsilon,\omega)\Phi(\epsilon)$. In the non-interacting case ($U=0$), we take $\rho(\epsilon,\omega)\sim \eta/((\omega+\mu-\epsilon)^2+\eta^2)$ with a small phenomenological broadening $\eta\simeq 0.001W$ mimicking the effect of impurity scattering. In this case, $\rho^2(\epsilon,\omega)$ is sharply peaked around $\omega=\epsilon-\mu$ and $\tilde{\Phi}(\epsilon,\omega)$ is effectively determined by the tight-binding transport DOS $\Phi(\epsilon)$. As a result, at low temperatures $A_1,A_0$ and hence $S$ is determined by the behavior of $\Phi(\omega+\mu)$ over an energy window $\sim k_\mathrm{B}T$ around the chemical potential ($\omega=0$). Hence the sign of $A_1$ or the Seebeck coefficient is negative (positive) depending on whether $\Phi(\omega+\mu)$ increases (decreases) with $\omega$, and $S\approx 0$ when $\omega=0$ is at the peak of $\Phi(\omega+\mu)$. As we show in Figs.\ref{fig:SelfEnergy_S}, for the interacting case treated within DMFT, due to the large and strongly temperature-dependent self-energy effects near the VHSs, $\tilde{\Phi}(\omega)$, unlike $\Phi(\omega+\mu)$, becomes very sharply peaked around the chemical potential and varies strongly over energy window $\sim k_\mathrm{B}T$. These lead to the non-standard sign and the violation of SMR over a range $-2\lesssim \nu\lesssim -1$ as shown in Fig.4 in the main text.

As shown in Fig. \ref{fig:SelfEnergy_S}, we also find strong particle-hole asymmetry in the self-energy and DOS even at low-energies near $\omega=0$ (chemical potential). The particle-hole asymmetry results from the particle-hole asymmetry of the non-interacting DOS at intermediate energies away from the VHSs, as evident in Fig.1d. The asymmetry is further enhanced by interaction. This asymmetry also influences the sign of the thermopower near the VHSs at low and intermediate temperatures. We also obtain a linear-$T$ resistivity (not shown) over an extended temperature range. This is, of course, a known outcome of DMFT for correlated systems \cite{Xu2013,Cha2019}. However, to obtain the detailed density dependence and the magnitude of resistivity one needs to consider the electron-phonon scattering and the effects of long-range impurities relevant at such low-densities near CNP \cite{Hwang2009,DasSarma2013}. The thermopower and the violation of SMR are not expected to be influenced significantly by electron-phonon and impurity scattering \cite{Jonson1980,Jonson1990}.

B.G. and P.S.M. contributed equally to this work.

\bibliography{Ref}